\documentclass[12pt,letterpaper]{article}
\usepackage{savesym}
\usepackage{pdflscape}
\usepackage{wrapfig}
\usepackage{amssymb}
\usepackage{amsthm}
\usepackage{amsmath}
\usepackage{graphicx}
\usepackage[usenames, table]{xcolor}
\usepackage{verbatim}
\usepackage{times}
\usepackage[titles]{tocloft}
\usepackage{enumitem}
\usepackage{bm}
\usepackage{sidecap}
\usepackage{epstopdf}
\usepackage{pdfpages}
\usepackage{hyphenat}
\usepackage[colorinlistoftodos]{todonotes}
\usepackage{soul}
\usepackage{multicol}
\usepackage{float}
\usepackage{multibib}
\usepackage{geometry}
\usepackage[small,compact]{titlesec}
\usepackage[normal,labelsep=colon,skip=2pt]{caption}
\usepackage{setspace}
\usepackage{subfig}
\usepackage{tocloft}
\usepackage{url}
\usepackage{floatflt}
\usepackage{array}
\usepackage{wasysym}
\usepackage{pgfgantt}
\usepackage[colorlinks=true,
            linkcolor=black,
            urlcolor=red,
            citecolor=blue]{hyperref}

\usepackage{cite}
\usepackage{lineno}
\usepackage[title,titletoc]{appendix}
\usepackage[affil-it]{authblk}

\newcommand*{\TitleFont}{%
      \usefont{\encodingdefault}{\rmdefault}{b}{n}%
      \fontsize{16}{20}%
      \selectfont}
\graphicspath{{./figs/}{../figs/}{../}}
\savesymbol{comment} 
\usepackage{changes}
\definechangesauthor[color=purple]{xl}
\definechangesauthor[color=orange]{hx}

\definecolor{lightblue}{rgb}{.90,.95,1}
\definecolor{darkgreen}{rgb}{0,.5,0.5}

\begin{document}

\title{\TitleFont LES wall modeling for heat transfer at high speeds}

\author[1]{Peng E S Chen}
\author[2]{Yu Lv}
\author[3]{Haosen H A Xu}
\author[4]{ Yipeng Shi}
\author[5]{Xiang I A Yang}

\date{\vspace{-5ex}}

\affil[1]{State Key Laboratory of Turbulence and Complex Systems, College of Engineering, Peking University, Beijing, China, 100871}
\affil[2]{Aerospace Engineering, Mississippi State University, Mississippi State, MS, USA, 39762}
\affil[3]{Mechanical Engineering, Pennsylvania State University, Pennsylvania, USA, 16802}
\affil[4]{State Key Laboratory of Turbulence and Complex Systems,College of Engineering, Peking University, Beijing, China, 100871}
\affil[5]{Mechanical Engineering, Pennsylvania State University, Pennsylvania, USA, 16802}

\maketitle

\noindent\makebox[\linewidth]{\rule{\linewidth}{0.6pt}}

\begin{abstract}
A practical application of universal wall scalings is near-wall turbulence modeling.
In this paper, we exploit temperature's semi-local scaling [Patel, Boersma, and Pecnik, {\it Scalar statistics in variable property turbulent channel flows}, Phys. Rev. Fluids, 2017, 2(8), 084604] and derive an eddy conductivity closure for wall-modeled large-eddy simulation of high-speed flows.
We show that while the semi-local scaling does not collapse high-speed direct numerical simulation (DNS) data, the resulting eddy conductivity and the wall model work fairly well.
The paper attempts to answer the following outstanding question: why the semi-local scaling fails but the resulting eddy conductivity works well.
We conduct DNSs of Couette flows at Mach numbers from $M=1.4$ to 6.
We add a source term in the energy equation to get a cold, a close-to-adiabatic wall, and a hot wall.
Detailed analysis of the flows' energy budgets shows that aerodynamic heating is the answer to our question: aerodynamic heating is not accounted for in Patel et al.'s semi-local scaling but is modeled in the equilibrium wall model.
We incorporate aerodynamic heating in semi-local scaling and show that the new scaling successfully collapses the high-speed DNS data.
We also show that incorporating aerodynamic heating or not, the semi-local scaling gives rise to the exact same eddy conductivity, thereby answering the outstanding question.
\end{abstract}

\noindent\makebox[\linewidth]{\rule{\linewidth}{0.6pt}}

\section{Introduction}
\label{sect:intro}

Engineering designs of aero-vehicles at high speeds are often aided by computational tools \cite{urzay2018supersonic, kiris2014lava, krist1998cfl3d, biedron2019fun3d}.
Compared to wall-resolving methods like direct numerical simulation (DNS) and wall-resolved large-eddy simulation (WRLES), wall modeled methods like hybrid RANS/LES method and wall-modeled large-eddy simulation (WMLES) are much more cost-effective at high Reynolds numbers \cite{choi2012grid,yang2021grid}.
Thanks to the recent developments in high-performance computing (HPC) \cite{kothe2018exascale}, the scale-resolving tool of WMLES has seen increased use for real-world engineering problems at low speeds \cite{park2016wall, bose2018wall}.
For example, WMLES of atmospheric boundary-layer flows are reported in Refs. \cite{giometto2016spatial,yang2016exponential, zhu2018turbulent,li2020revisiting,zhou2020wall,ge2021large}, Xu et al. reported WMLES of a compressor's return channel \cite{xu2021comparative}, Park et al. presented WMLES results for the NASA common research model \cite{lehmkuhl2016flow}, Ghate et al. compared WMLES, DDES, and RANS for the NASA juncture flow \cite{ghate2020scale}, Huang, Xu, and coauthors reported WMLES of flow around complex and moving boundaries \cite{ma2019dynamic, ma2021hybrid, wang2021off}, to name a few.
Because of a lack of model validation, there are fewer uses of WMLES for high-speed flows, see, e.g., Refs. \cite{iyer2018large, bermejo2014confinement}, and even fewer for high-speed heat transfer, see, e.g., Refs. \cite{yang2018aerodynamic,fu2021shock}.
Two open questions in the field are:
How to close the eddy conductivity for high-speed flows?
How does the wall model do, and why does it work/not work?
The objective of this work is to answer the above two questions.
In the following, we briefly review the existing literature on this topic.
We would also show some data to better motivate the present work.

\begin{figure}
\centering
\includegraphics[width=0.45\textwidth]{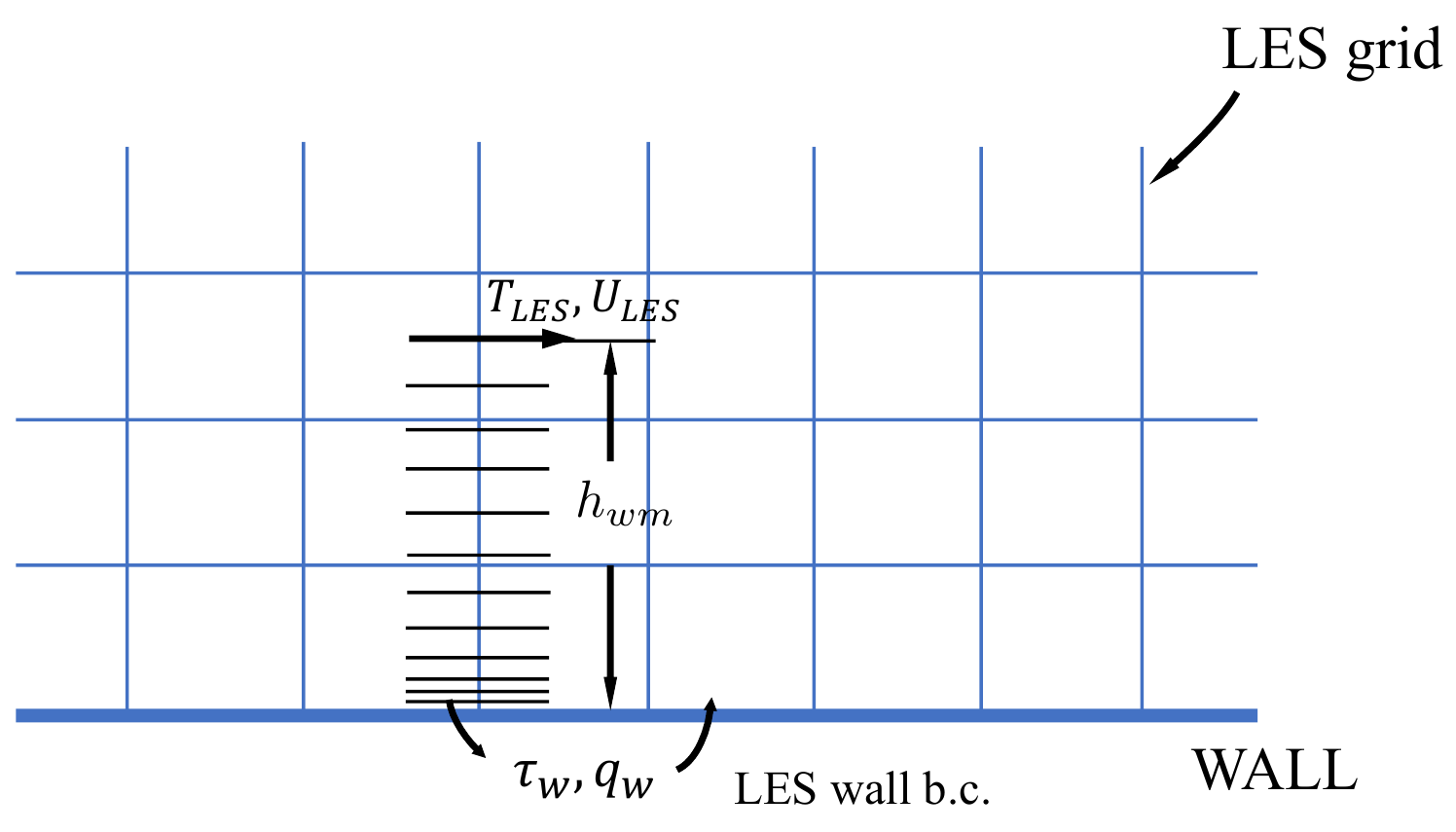}
\caption{A sketch of WMLES.
The wall model is a stress-based one.
}
\label{fig:WM-sketch}
\end{figure}

Figure \ref{fig:WM-sketch} is a sketch of a WMLES.
The wall model models the wall-shear stress and the wall heat transfer rate.
The most extensively used wall model is the equilibrium wall model (EWM) \cite{kawai2012wall}, which reads
\begin{subequations}\label{eq:WM}
\begin{linenomath*}\begin{equation}
\small
    \dfrac{d}{d y} \left[ (\mu + \mu_{t,wm}) \frac{d u_\parallel}{d y}\right] = 0,
    \label{eq:WM-U}
\end{equation}\end{linenomath*}
\begin{linenomath*}\begin{equation}
\small
    \dfrac{d}{ d y} \left[ (\mu + \mu_{t,wm})u_\parallel \dfrac{d u_\parallel}{d y} + c_p \left( \frac{\mu}{Pr} + \frac{\mu_{t,wm}}{Pr_{t,wm}}\right) \dfrac{d T}{d y}\right]=0,
    \label{eq:WM-T}
\end{equation}\end{linenomath*}
\end{subequations}
where $\mu$ is the molecular dynamic viscosity, $\mu_{t,wm}$ is wall model's eddy viscosity, $u_{||}$ is the wall-parallel velocity, $c_p$ is the specific heat, $Pr$ is the molecular Prandtl number, $Pr_t$ is the turbulent Prandtl number ($\mu_{t,wm}/Pr_t$ is the eddy conductivity), and $T$ is the temperature.
The eddy viscosity and the eddy conductivity in the wall model equations must be closed, and a commonly-used closure is 
\begin{linenomath*}\begin{equation}
    \mu_{t,wm} = \sqrt{\rho \tau_w}\kappa y D,
    \label{eq:WM-mut}
\end{equation}\end{linenomath*}
and $Pr_t=0.9$.
Here, $\rho$ is the fluid density, $\tau_w$ is the wall-shear stress, $\kappa=0.41$ is the von Karman constant, and 
\begin{linenomath*}\begin{equation}
    D=\left[1-\exp\left(-y^+/A^+\right)\right]^2
    \label{eq:wm-D}
\end{equation}\end{linenomath*}
is the van Driest damping function with $A^+=17$ being the van-Driest damping function constant.
Equations \eqref{eq:WM-U} and \eqref{eq:WM-T} are solved on a 1D grid between the wall $y=0$ and a matching location $y=h_{wm}$ with the following matching condition at $y=h_{wm}$
\begin{linenomath*}\begin{equation}
    u_{||}(y=h_{wm})=U_{LES},~~T(y=h_{wm})=T_{LES},
    \label{eq:wm-bc}
\end{equation}\end{linenomath*}
where the subscript $LES$ denotes large-eddy simulation (LES) quantities.
Although there are complications like log-layer mismatch \cite{kawai2012wall,yang2017log} and near-wall resolution \cite{iyer2018large,xu2021assessing} when applying the EWM in a WMLES, the model gives accurate estimates of the mean flow if a constant stress layer exists. 
In fact, at low speeds, the energy equation Eq. \eqref{eq:WM-T} degenerates to $T=$Const, and integrating Eqs. \eqref{eq:WM-U} leads to
$
    U^+=\log(y^+)/\kappa+B,
$
i.e., the well-established logarithmic law in the constant stress layer \cite{xu2018fractality, marusic2013logarithmic}, which is a direct {\it a priori} validation of the EWM at low speeds.
Here, the superscript $+$ denotes normalization by the wall units, and $B\approx 5.2$ is a constant.
The largest uncertainty comes from the von K{\'a}rm{\'a}n constant, which is responsible for a 5\% error.
{\it A priori} validation of the EWM at high speeds is much more difficult because there is no well-established universal law of the wall.
Considering that one relies on the law of the wall to close the wall model equations, the lack of a well-established universal law of the wall partly explains why WMLES is considered to be less reliable for flows at high speeds than at low speeds.

\begin{figure*}
\centering
\includegraphics[width=0.9\textwidth]{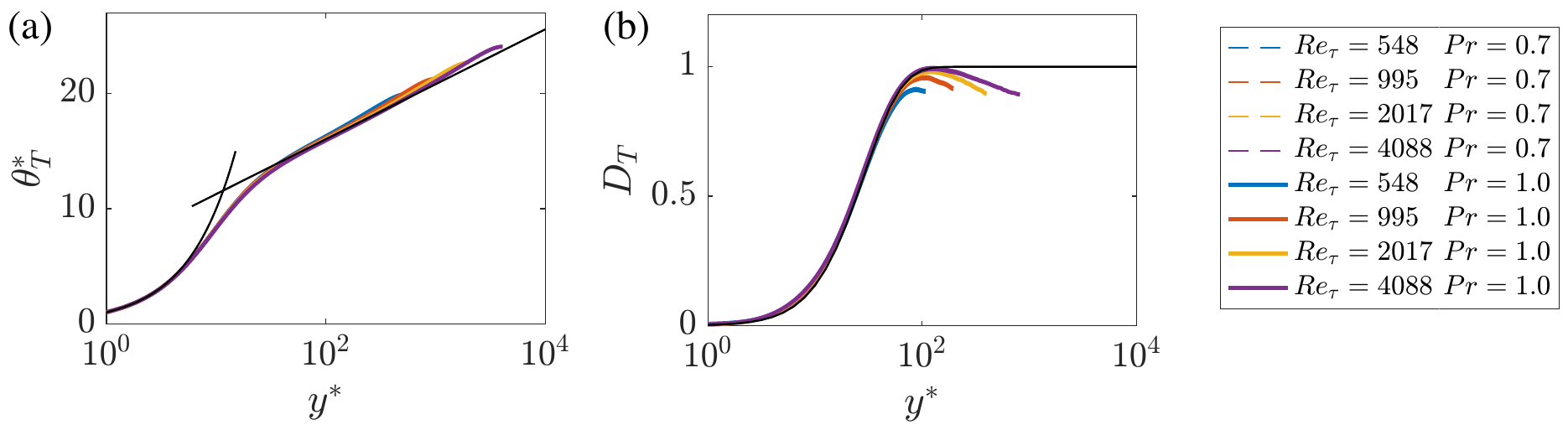}
\caption{(a) $\theta_T$ as a function of $y^*$ for incompressible channel flow at $Re_\tau=548$, 995, 2017, and 4088 and for $Pr=0.7$ and 1.0 \cite{pirozzoli2016passive}.
The two thin black lines are $\theta_T=y^*$ and $\theta_T=1/\kappa_T\log(y^*)+B'$.
For incompressible flows, $\theta^*=\theta/\theta_\tau$, $y^*=y^+$.
(b) $D_T=((d\theta_T/dy^*)^{-1}-1)/(\kappa y^*)$ for $y/\delta<0.2$, i.e., below the logarithmic layer.
The thin black line corresponds to $D_T=(1-\exp(-y/A_T))^2$, where $A_T=20$.
}
\label{fig:semi-incomp}
\end{figure*}

\begin{figure}
\centering
\includegraphics[width=0.32\textwidth]{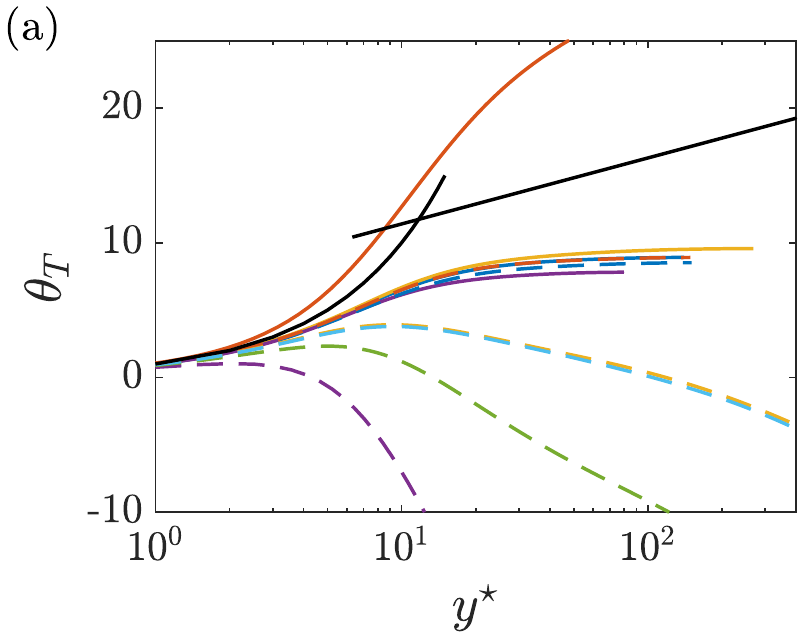}~~~\includegraphics[width=0.32\textwidth]{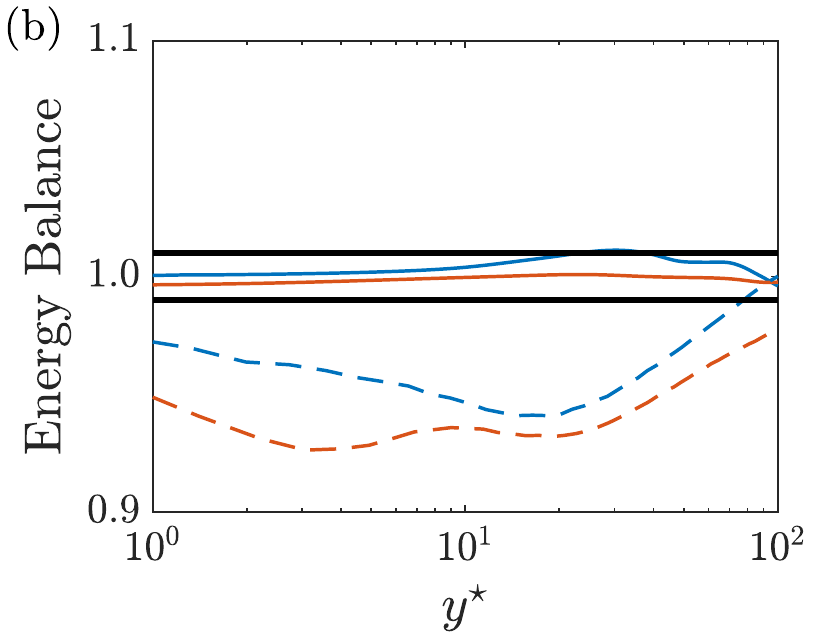}\\
\includegraphics[width=0.28\textwidth]{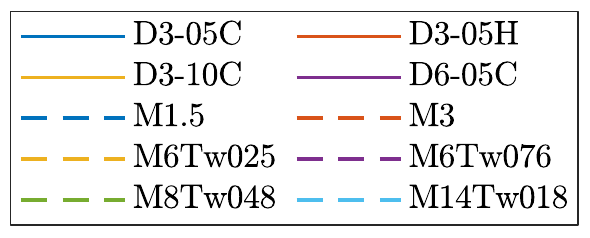}
\caption{(a) $\theta_T$ as a function of $y^*$ for our DNSs: D3-05C, D3-05H, D3-10C, and D6-05C,
Zhang et al's DNSs: M6Tw025, M6Tw076, M8Tw048, and M14Tw018 \cite{zhang2018direct}, and Coleman et al's DNSs: M1.5 and M3.
Table \ref{tab:DNS} tabulates our DNSs' details.
The nomenclature of Zhang et al.'s DNSs is as follows: M[A]Tw[B], where A is the freestream Mach number, and $B=T_w/T_r\times 100$, and $T_r$ is the recovery temperature.
The nomenclature of Coleman et al.'s DNSs is as follows: M[A], where A is the bulk Mach number \cite{coleman1995numerical}.
The two thin black lines are $\theta_T=y^*$ and $\theta_T=1/\kappa_T\log(y^*)+B'$.
(b) The normalized sum of the terms in the energy equation of our DNSs and Coleman et al.'s DNSs. 
The sum should be 1 if the data is statistically converged.
The error in the momentum equation is usually less than 1\% \cite{yang2021resolution,oliver2014estimating}. 
The two black lines enclose the region $1\pm 0.01$.
}
\label{fig:semi-comp}
\end{figure}

Finding a universal law of the wall for the fluid velocity and the fluid temperature in high-speed boundary layer flows is a continuous effort in fundamental flow physics research.
The research on this topic dates back to van Driest \cite{van1956turbulent}, Morkovin \cite{morkovin1962effects}, and Huang et al. \cite{huang1995compressible, huang1994van}.
The more recent developments are by Trettel \& Larsson \cite{trettel2016mean} and Patel et al. \cite{patel2017scalar}.
Patel et al. argue that turbulent transport and viscous diffusion are the dominant terms in the Reynolds averaged NS equations. 
Their hypothesis gives rise to the following semi-local scaling 
\begin{linenomath*}\begin{align}
   \frac{dU^*}{dU}&=\left(1+\frac{y}{Re_\tau^*}\frac{dRe_\tau^*}{dy}\right)\sqrt{\frac{\rho}{\rho_w}}\big/u_\tau, \nonumber \\
    U^*&\equiv y^*, ~~\text{in the viscous sublayer,} \label{eq:semi-U}\\
    U^*&\equiv\frac{1}{\kappa}\log(y^*)+B, ~~\text{in the log layer;} \nonumber
\end{align}\end{linenomath*}
and 
\begin{linenomath*}
\begin{subequations}\label{eq:semi-T}
\begin{equation}
\begin{split}
    \frac{d\theta^*}{d\theta}&=\left(1+\frac{y}{Re_\tau^*}\frac{dRe_\tau^*}{dy}\right)\sqrt{\frac{\rho}{\rho_w}}\big/\theta_\tau,  \\
    \theta^*&\equiv Pr^*~y^*, ~~\text{in the viscous sublayer,} \\
    \theta^*&\equiv\frac{1}{\kappa_T}\log(y^*)+f(Pr^*), ~~\text{in the log layer;} 
\end{split}
    \label{eq:semi-T1}
\end{equation}
\begin{equation}
\begin{split}
    \frac{d\theta_T}{d\theta^*}&=\frac{1/Pr^*+\alpha_t/\mu}{1+\alpha_t/\mu},\\
    \theta_T&\equiv y^*, ~~\text{in the viscous sublayer,} \\
    \theta_T&\equiv\frac{1}{\kappa_T}\log(y^*)+B', ~~\text{in the log layer.}
\end{split}
    \label{eq:semi-T2}
\end{equation}
\end{subequations}
\end{linenomath*}
The above semi-local scaling maps the velocity $U$ and the temperature $T$ (or $T-T_w$) to $U^*$ and $\theta_T$, both of which are universal functions of $y^* =\sqrt{{\rho}\tau_w}y/\mu$.
Here, $1/\kappa_T\approx 2.12$ \cite{kader1981temperature}, $f(Pr^*)$ is a function of $Pr^*$, $Pr^* = \mu/\lambda$, $B'\approx 6.5$ is a constant, ${\theta}={T}-T_w$, $\theta_\tau = q_w/(\rho_w c_{pw} u_\tau)$, $q_w$ is the wall heat flux, $u_\tau=\sqrt{\tau_w/\rho_w}$ is the friction velocity,  $Re_\tau^* =  \sqrt{{\rho \rho_w}} u_\tau/{\mu}$, $\lambda$ is the molecular conductivity, and $\alpha_t=Pr_t\mu_t$ is the eddy conductivity.
The subscript $w$ denotes quantities evaluated at the wall, and the superscript $*$ denotes normalization by local quantities (as opposed to wall quantities).
All quantities in Eqs. \eqref{eq:semi-U} and \eqref{eq:semi-T} are Reynolds averaged.
While the velocity scaling has been somewhat successful \cite{modesti2016reynolds,pecnik2017scaling,zhang2018direct,yang2018semi,yao_supersonic_2019}, the temperature (scalar) scaling has received much less attention.
Figure \ref{fig:semi-incomp} (a) and figure \ref{fig:semi-comp} (a) compare $\theta_T$ to the expected universal scaling in incompressible \cite{pirozzoli2016passive} and compressible flows.
Evaluating $\theta_T$ proves to be not possible for many of the online DNS databases \cite{modesti2016reynolds,volpiani2018effects}:
First, the evaluation of $\theta_T$ involves the evaluation of the eddy conductivity, $\alpha_T=-\rho\overline{\theta''v''}/d\theta/dy$, but 
{$\overline{\theta''v''}$ is not available.}
Second, the data are not statistically converged. Figure \ref{fig:semi-comp} (b) shows the sum of the terms in the energy equation for the DNSs in Ref. \cite{coleman1995numerical}.
We see a close to 7\% convergence error.
While Coleman et al. have reported all the terms in the energy equation, many of the databases do not, and we cannot directly evaluate their statistical convergence.
We would revisit this point in sections \ref{sect:results} and \ref{sect:discussion}.
There, we shall see that the thermal field is more challenging to compute than the momentum field.
In all, we conclude from figures \ref{fig:semi-incomp} (a) and \ref{fig:semi-comp} (a) that while the semi-local scaling collapses the low speed data, it fails to do so for high-speed flows.

In addition to collapsing data, a transformation that maps $U$ and $T$ to a universal scaling gives closure models for the eddy viscosity/conductivity.
Yang and Lv \cite{yang2018semi} exploited the semi-local scaling and concluded that the damping function must be
\begin{linenomath*}\begin{equation}
    D_{sl}=\left[1-\exp(-y^*/A)\right]^2.
    \label{eq:wm-Dsl}
\end{equation}\end{linenomath*}
The use of semi-local scaling in the damping function was also reported in Ref. \cite{bocquet2012compressible}---although, there, its use is empirical.
In a recent work, \cite{iyer2019analysis} proposed a mixed scaling, which gives more accurate results than Eq. \eqref{eq:wm-Dsl} in a WMLES context.
To date, research on this topic is limited to the momentum field and is yet to extend to the thermal field.

In section \ref{sect:derivation}, we exploit Eq. \eqref{eq:semi-T} and close the energy equation.
Given the poor performance of the semi-local scaling in figures \ref{fig:semi-incomp} (a) and \ref{fig:semi-comp} (a), it would seem that relying on the scaling in Eq. \eqref{eq:semi-T} for closure models is an ill-conceived idea.
The results, however, are surprising.
In fact, Iyer \& Malik \cite{iyer2019analysis} assessed a similar model and concluded that the EWM works well for high speed flows: the error is less than about 5\% if the LES/wall-model matching location is not above about 10\% of the half channel height and if one employs the semi-local scaling.
The outstanding question is why the semi-local scaling fails but the resulting wall model works.

The rest of the paper is organized as follows.
In section \ref{sect:derivation}, we exploit the semi-local scaling to close the energy equation.
We conduct DNSs and WMLESs of turbulent Couette flows for {\it a priori} and {\it a posteriori} analyses.
Details of the DNSs and WMLESs are shown in section \ref{sect:setup}.
In section \ref{sect:results}, we show that the aerodynamic heating term $(\mu+\mu_{t,wm})u_{||}du_{||}/dy$ is the answer to our question.
Incorporating the aerodynamic heating term in the semi-local scaling, we show in section \ref{sect:discussion} that the semi-local scaling collapses the high-speed data.
Finally, conclusions are given in section \ref{sect:conclusions}.

\section{Semi-local scaling and eddy conductivity}
\label{sect:derivation}

In this section, we exploit the semi-local scaling in Eq. \eqref{eq:semi-T} and model the eddy conductivity in a constant stress layer.
The basic logic is to accept all the assumptions in Ref \cite{patel2017scalar} and derive the eddy conductivity.
We will revisit the validity of some assumptions in section \ref{sect:discussion}.
Following Patel et al. \cite{patel2017scalar}, we neglect aerodynamic heating and mean flow convection.
The eddy conductivity is
\begin{linenomath*}\begin{equation}
    \alpha_t=-\rho \overline{v''\theta''}\big/\frac{d\theta}{dy}.
    \label{eq:alphat}
\end{equation}\end{linenomath*}
where $-\rho \overline{v''\theta''}$ is the turbulent heat flux, and $\overline{\cdot}$ denotes Reynolds averaged.
The energy equation in the constant stress layer reads \cite{patel2017scalar}:
\begin{linenomath*}\begin{equation}
    -\rho \overline{v''\theta''}+\frac{\mu}{Pr^*} \frac{d\theta}{dy}=q_w/c_{pw},
    \label{eq:energy-Patel}
\end{equation}\end{linenomath*}
and therefore 
\begin{linenomath*}\begin{equation}
    -\rho \overline{v''\theta''}=q_w/c_{pw}-\frac{\mu}{Pr^*} \frac{d\theta}{dy}=\rho_w u_\tau \theta_\tau-\frac{\mu}{Pr^*} \frac{d\theta}{dy}.
    \label{eq:rhovT}
\end{equation}\end{linenomath*}
Per Eq. \eqref{eq:semi-T}, we have
\begin{linenomath*}\begin{equation}
\begin{split}
    \frac{d\theta}{dy}
    &=\frac{d\theta}{d\theta^*}\frac{d\theta^*}{d\theta_T}\frac{d\theta_T}{dy^*}\frac{dy^*}{dy}\\
    &=\frac{\theta_\tau}{\left(1+\dfrac{y}{Re_\tau^*}\dfrac{dRe_\tau^*}{dy}\right)\sqrt{\dfrac{\rho}{\rho_w}}} \times\frac{1+\alpha_t/\mu}{1/Pr^*+\alpha_t/\mu}\times\frac{d\theta_T}{dy^*}\times\left(1+\frac{y}{Re_\tau^*}\frac{dRe_\tau^*}{dy}\right)\frac{\sqrt{{\rho}\tau_w}}{{\mu}}\\
    &=\rho_wu_\tau\theta_\tau\frac{d\theta_T}{dy^*}\frac{1+\alpha_t/\mu}{1/Pr^*+\alpha_t/\mu}\frac{1}{\mu}.
\end{split}
\label{eq:dthtdy}
\end{equation}\end{linenomath*}
In Eq. \eqref{eq:dthtdy}, $d\theta_T/dy^*$ is (supposably) a universal function of $y^*$, and according to Eq. \eqref{eq:semi-T2}
\begin{linenomath*}\begin{equation}
\begin{split}
    \frac{d\theta_T}{dy^*}&=1 ~~\text{in the viscous sublayer,}\\
    \frac{d\theta_T}{dy^*}&=\frac{1}{\kappa_T y^*} ~~\text{in the logarithmic layer.}
\end{split}
\end{equation}\end{linenomath*}
We make an analogy to the velocity scaling and define a damping function $D_T(y^*)$ such that 
\begin{linenomath*}\begin{equation}
    \frac{d\theta_T}{dy^*}=\frac{1}{1+\kappa_Ty^*D_T(y^*)}.
    \label{eq:DT}
\end{equation}\end{linenomath*}
This damping function $D_T$ is 0 when $y^*\to 0$ and $1$ when $y^*\gg 1$.
Thanks to the universality of $\theta_T$, we can measured $D_T$ in a incompressible flow and expect the measured $D_T$ to work at high speeds.
Figure \ref{fig:semi-incomp} (b) shows $D_T$, i.e., $D_T=[(d\theta_T/dy^*)^{-1}-1]/(\kappa_T y^*)$, as a function of $y^*$ in flows at Reynolds numbers from $Re_\tau=548$ to $4088$.
There is an apparent Reynolds number effect, but at a sufficiently high Reynolds number and for $y\ll h$,
\begin{linenomath*}\begin{equation}
D_T(y^*)=\left[1-\exp\left(-y^*/A_T\right)\right]^2, ~~~~A_T=20,
\label{eq:DT2}
\end{equation}\end{linenomath*}
seems to be a good working approximation.
Here, we emphasize that $y$'s scaling in the damping function $D_T$ is solely controlled by $\theta_T$'s scaling.
While one may employ other scalings for $y$ in the damping function, and it may give more accurate results, such efforts are necessarily empirical \cite{iyer2019analysis, volpiani2020data}.
(This is by no means a criticism of the previous work: LES wall modeling is in and of itself empirical.)

\begin{figure}
\centering
\includegraphics[width=0.32\textwidth]{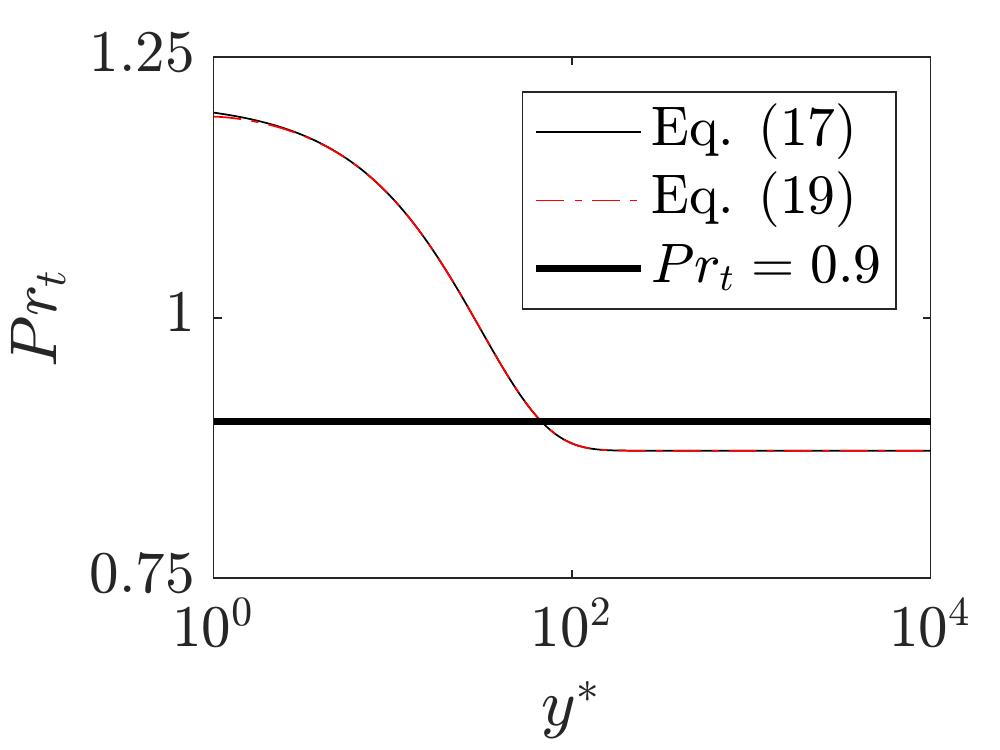}
\caption{Turbulent Prandtl number model.  $\epsilon=10^{-6}$ in Eq. \eqref{eq:WM-Prt-implement}.
}
\label{fig:Prt}
\end{figure}

Equations \eqref{eq:rhovT}, \eqref{eq:dthtdy}, and \eqref{eq:DT} give
\begin{linenomath*}\begin{equation}
    \alpha_t=\frac{\left( 1+\kappa_T y^*D_T\right)\left({\mu}/{Pr^*} + \alpha_t\right)}{1+{\alpha_t}/{\mu}} -\frac{\mu}{Pr^*}.
    \label{eq:alphat2}
\end{equation}\end{linenomath*}
Equation \eqref{eq:alphat2} is a quadratic equation of $\alpha_t$, whose only physical solution is
\begin{linenomath*}\begin{equation}
    \alpha_t/\mu=\kappa_T y^* D_T(y^*).
    \label{eq:semi-alphat}
\end{equation}\end{linenomath*}
Alternatively,
\begin{linenomath*}\begin{equation}
    Pr_t=\frac{\mu_t}{\alpha_t}=\frac{\kappa}{\kappa_T}\frac{D_{sl}}{D_T}=\frac{\kappa}{\kappa_T}  \left[\frac{1-\exp(-y^*/A)}{1-\exp(-y^*/A_T)}\right]^2
    \label{eq:WM-Prt}
\end{equation}\end{linenomath*}
Although the denominator $[1-\exp(-y^*/A_T)]^2=0$ when $y^*=0$, Eq. \eqref{eq:WM-Prt} is non-singular at $y^*=0$:
\begin{linenomath*}\begin{equation}
    \lim_{y^*\to 0}Pr_t=\frac{\kappa}{\kappa_T}\left[\frac{{y^*}/{A}}{y^*/A_T}\right]^2=\frac{\kappa}{\kappa_T}\left(\frac{A_T}{A}\right)^2.
\end{equation}\end{linenomath*}
But one may still encounter numerical difficulty if (s)he directly implements Eq. \eqref{eq:WM-Prt}, in which case, the following formulation should be used
\begin{linenomath*}\begin{equation}
    Pr_{t,\epsilon}=\frac{\mu_t}{\alpha_t}=\frac{\kappa}{\kappa_T}  \frac{\left[1-\exp(-y^*/A)\right]^2+A_T^2\epsilon}{\left[1-\exp(-y^*/A_T)\right]^2+A^2\epsilon},
    \label{eq:WM-Prt-implement}
\end{equation}\end{linenomath*}
where $\epsilon$ is a small number.
Figure \ref{fig:Prt} shows the turbulent Prandtl number in Eqs. \eqref{eq:WM-Prt} and \eqref{eq:WM-Prt-implement} as well as that in the conventional viscous scaled WM, i.e., $Pr_t=0.9$.
The turbulent Prandtl in Eq. \eqref{eq:WM-Prt} number is a decreasing function of $y^*$ and asymptotes to $\kappa/\kappa_T=0.85$ for sufficiently large $y^*$, which are consistent with Refs \cite{kays1994turbulent,weigand1997extended}.
The asymptotic value at $y^*=0$, i.e., $Pr_t=1.21$, does not completely agree with the recent high-speed DNSs, see, e.g., Refs \cite{priebe2021turbulence, fu2021shock, di2021direct}---however, we emphasize that logic here is to accept all assumptions in Ref. \cite{patel2017scalar}.

Equation \eqref{eq:WM}, \eqref{eq:WM-mut}, \eqref{eq:wm-Dsl}, and \eqref{eq:WM-Prt} constitute our wall model.
Equation \eqref{eq:WM-Prt} is the contribution of this work.
The wall model closures, i.e., the eddy viscosity and the eddy conductivity (the turbulent Prandtl number), are direct consequences of the semi-local scaling.

\section{Computational setup}
\label{sect:setup}

\begin{figure}
\centering
\includegraphics[width=0.5\textwidth]{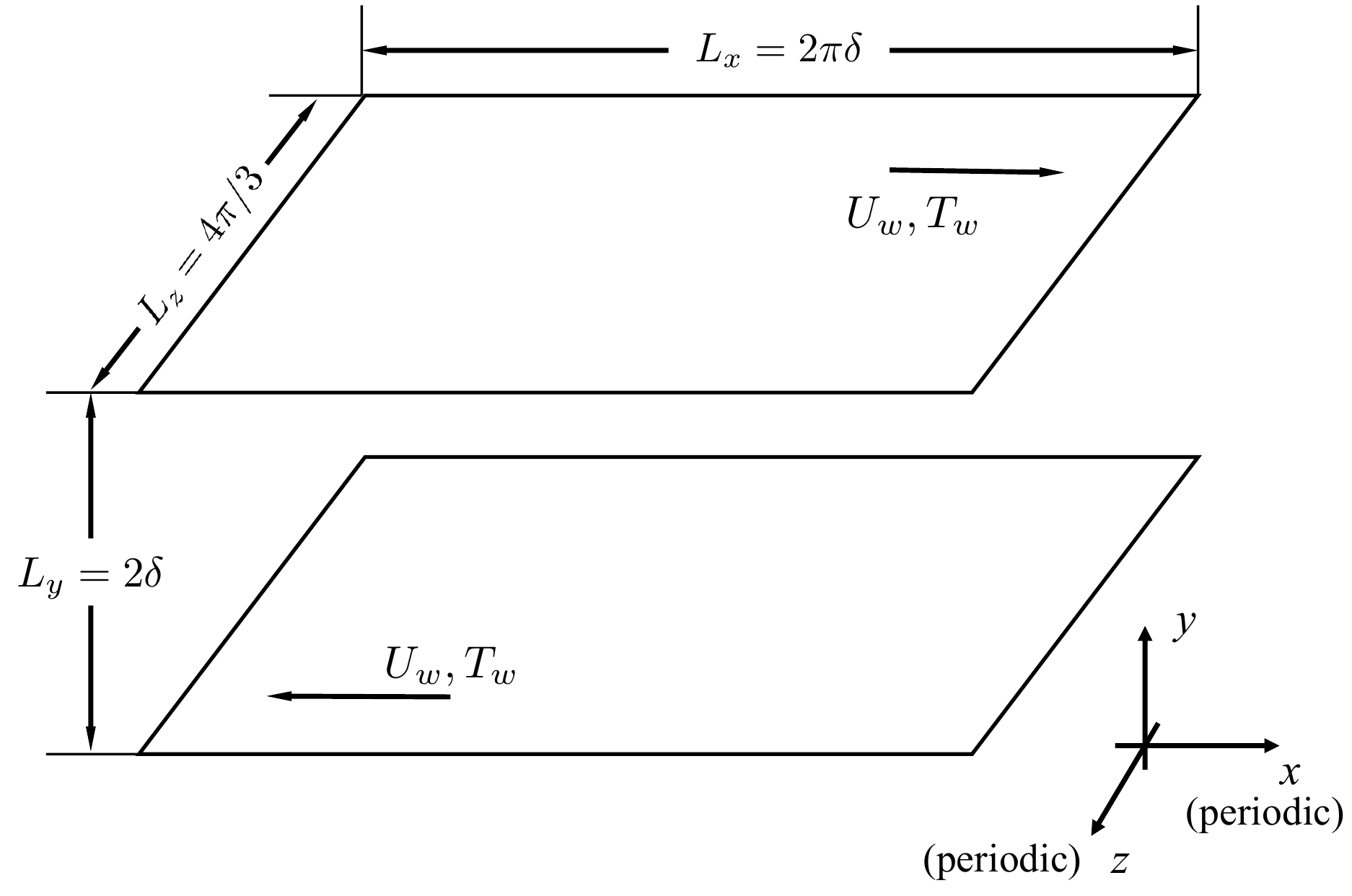}
\caption{A sketch of the flow configuration.
$x$, $y$, and $z$ are the streamwise, wall-normal, and spanwise coordinates. 
The flow is periodic in the streamwise and the spanwise directions.
The two walls are at the same temperature $T_w$.
The wall velocity is $U_w$.
$\delta$ is the half channel height.
}
\label{fig:domain}
\end{figure}

Rather than entirely relying on the online databases, we will also conduct our own DNSs. 
The online DNS databases do not serve the purpose of this work because of a lack of statistical convergence and a lack of some flow statistics, which we will discuss in further detail in section \ref{sect:discussion}.
Figure \ref{fig:domain} shows a sketch of the flow configuration.
The mean pressure gradient is 0 in Couette flow, and the logarithmic layer extends closer to the channel centerline than, e.g., in a channel flow \cite{lee2018extreme}.
The absence of an apparent wake layer allows a direct comparison between the flow and the EWM \cite{yang2018semi,yang2018aerodynamic}.
In the next two subsections, we present the detailed setups of our DNSs and LESs.

\subsection{DNS}

\begin{table}
    \caption{\label{tab:DNS}DNS details. 
    $\Delta x$, $\Delta y$ and $\Delta z$ are the grid spacings in streamwise, wall-normal, and spanwise directions. 
    We show both the grid resolution at the wall (the first number), i.e.,  $\Delta x^+_w$, $\Delta y^+_w$, and $\Delta z^+_w$, and the grid resolution at the channel centerline, i.e., $\Delta x^*_c$, $\Delta y^*_c$, and $\Delta z^*_c$ (the second number). 
    $M_{\rm ref}=U_w/a_{\rm ref}$, where
    $a_{\rm ref}=\sqrt{\gamma R T_{\rm ref}}$ is the speed of sound at the reference temperature, and $\gamma=1.4$ is the specific heat ratio.
    M is the wall Mach number.
    The molecular Prandtl number is $Pr\equiv Pr^*=0.7$.
    $Re_b=\rho_b U_w\delta/\mu_{\rm ref}$ is the bulk Reynolds number, where $\rho_b$ is the bulk density.
    The computational domain size is $L_x\times L_y\times L_z=2\pi\delta\times 2\delta\times 4/3 \pi\delta$.
    $\varphi$ is a heat source.
    $b_g$ controls the grid stretching and is defined in Eq. \eqref{eq:DNSgrid}.
    }
    \vspace{2mm}
    \begin{tabular}{|c|>{\centering\arraybackslash}m{0.04\textwidth}|
    >{\centering\arraybackslash}m{0.03\textwidth}|
    >{\centering\arraybackslash}m{0.07\textwidth}|
    >{\centering\arraybackslash}m{0.08\textwidth}|
    >{\centering\arraybackslash}m{0.09\textwidth}|
    >{\centering\arraybackslash}m{0.23\textwidth}|
    >{\centering\arraybackslash}m{0.04\textwidth}|
    >{\centering\arraybackslash}m{0.04\textwidth}|
    >{\centering\arraybackslash}m{0.04\textwidth}|}
    \hline
        Case & $M_{\rm ref}$ & $M$ & $Re_b$ & $T_w/T_{\rm ref}$ & $\varphi \delta/\rho_b U_w^3$ & $N_x \times (N_y, b_g) \times N_z$ & $\Delta x$ & $\Delta y$ & $\Delta z$  \\ \hline
         D3-05C & 3 & 3   & 5000  & 1.0 & 0       & $256 \times (257, 2.0) \times 256$ & 9.8, 3.5 & 0.47, 2.3 & 6.56, 2.3\\\hline
         D3-05A & 3 & 1.9 & 5000  & 2.5 & -0.0029 & $256 \times (161, 2.0) \times 256$ & 2.3, 6.7 & 0.18, 7.1 & 1.54, 4.5\\\hline
         D3-05H & 3 & 1.4 & 5000  & 4.5 & -0.0050 & $256 \times (161,2.0) \times 256$ & 1.7, 5.0  & 0.13, 5.2 & 1.13, 3.3\\\hline
         D3-10C & 3 & 3   & 10000 & 1.0 & 0       & $512 \times (241,2.3) \times 512$ & 9.4, 3.3 & 0.59, 5.26 & 6.30, 2.2\\\hline
         D3-10A & 3 & 1.8 & 10000 & 2.8 & -0.0025 & $416 \times (209,1.0) \times 400$ & 2.5, 6.8  & 0.85, 5.7 & 1.68, 4.7\\\hline
         D6-05C & 6 & 6   & 5000  & 1.0 & 0       & $384 \times (385,2.0) \times 512$ & 11, 1.3  & 0.53, 0.86& 8.4, 0.98\\\hline
    \end{tabular}
\end{table}

We use the in-house high-order finite-difference code Hoam-OpenCFD for our DNSs.
The code solves the full compressible Navier-Stokes equations.
The working fluid is an ideal gas.
The molecular viscosity varies with the temperature according to the Sutherland's law
\begin{linenomath*}\begin{equation}
    \frac{\mu}{\mu_{\rm ref}}=\left(\frac{T}{T_{\rm ref}}\right)^{3/2}\frac{T_{\rm ref}+T_s}{T+T_s}
\end{equation}\end{linenomath*}
where $T_s=110.4K$ and $T_{\rm ref}=288.15K$ are the Sutherland temperature and the reference temperature, respectively.
Further details of the code can be found in Ref. \cite{li2006direct} and the references cited therein.
The size of the computational domain is $L_x\times L_y\times L_z=2\pi\times 2 \times 4\pi/3 (\delta)$ \cite{lozano2014effect}.
We use a uniform grid in both the streamwise and the spanwise directions.
The grid in the wall-normal direction follows
\begin{linenomath*}\begin{equation}
    {y_j}/{\delta}=\tanh{\left[b_g\left(2\frac{j-1}{N_y-1}-1\right)\right]}\big/\tanh{(b_g)},
    \label{eq:DNSgrid}
\end{equation}\end{linenomath*}
where $j=1$, 2, ..., $N_y$, and $N_y$ is the number of wall-normal grid points, $b_g$ controls the grid stretching.
The grid is such that the resolution is comparable/finer than the previous DNSs  \cite{pirozzoli2004direct,pirozzoli2011direct, pirozzoli2011turbulence, modesti2015high,volpiani2018effects}.
In the absence of any heat source, the viscous heating is balanced by the heat transfer at the wall, and the two walls are cold.
We include a negative heat source such that the wall is hot, nearly adiabatic, and cold.
Table \ref{tab:DNS} shows the other details of our DNSs.
The nomenclature is D$[U_w/a_{ref}]$-$[Re_b/1000]$C/A/H.
The first letter is for ``DNS''.
The last letter ``C'' is for ``cold wall'', ``A'' is for ``nearly adiabatic wall'', and ``H'' is for ``hot wall''.

\begin{figure}
\centering
\includegraphics[width=0.32\textwidth]{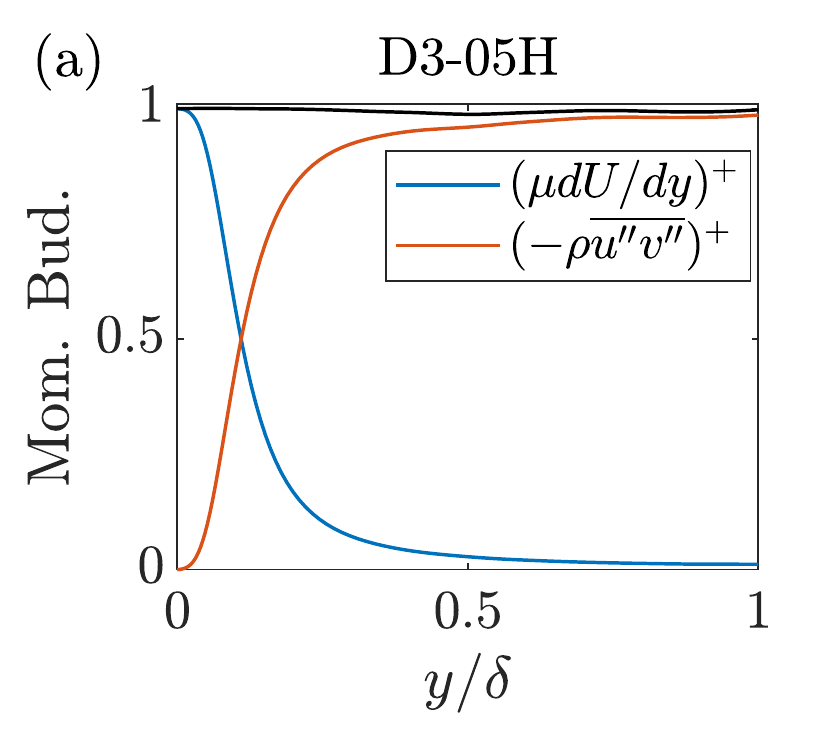}
\includegraphics[width=0.32\textwidth]{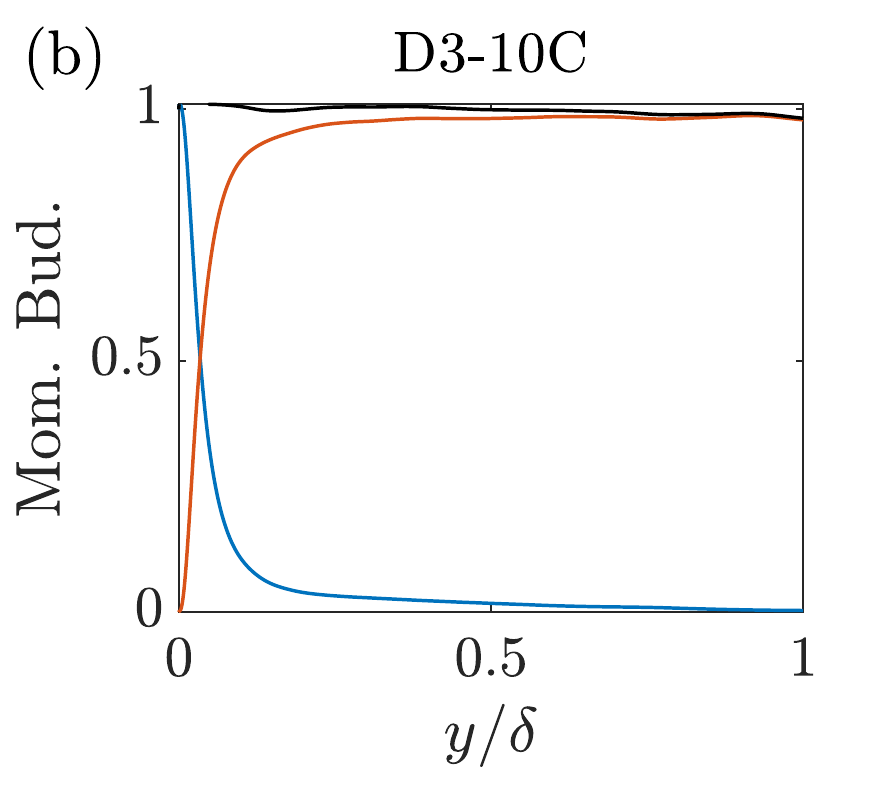}
\caption{Viscous and turbulent momentum fluxes as a function of $y/\delta$ in (a) Case D3-05H and (b) Case D3-10C.
``Mom. Bud.'' is short for ``momentum budget''.
The thin black line is the sum of the two flux.
}
\label{fig:mom-bud}
\end{figure}

Following Ref. \cite{oliver2014estimating}, we check the convergence of our DNSs by examining the momentum budget.
In a Couette flow, $-\rho \overline{u''v''}+\mu dU/dy=\tau_w$.
Figure \ref{fig:mom-bud} (a, b) shows the turbulent flux and the viscous flux as a function of the wall-normal coordinate in cases D3-10C and D3-05H.
There is no heat source in D3-10C, and D3-05H has a non-zero heat source. 
Results of other DNSs are similar and are not shown here for brevity.
The error in the total flux is about 1\% and is comparable to the previous DNSs \cite{lee2015direct,pirozzoli2016passive}.
We may also check the convergence by examining the energy (the total energy, not the turbulent kinetic energy) budget.
The results are shown in figure \ref{fig:ener-bud} in section \ref{sect:discussion}.
There, the error is also less than 1\%.
Table \ref{tab:DNS2} shows the further details of our DNSs.
We have reported the Brinkman number, $Br=q_w/(\tau_w U_w)=$ heat transfer/aerodynamic heating. 
For the $C$ cases, i.e., D3-05C, D3-10C, and D6-05C, the heat source is 0 and energy conservation dictates that $q_w=U_w\tau_w$ and $Br=1.0$.
For the $A$ and the $H$ cases, where there is a negative heat source, the energy conservation is such that $q_w=\tau_w U_w-\left|\varphi \delta\right|$, and $\left|Br\right|<1$.
Hence, aerodynamic heating is non-negligible in all our DNSs.

\begin{table}
    \caption{\label{tab:DNS2}Further DNS details. $Re_\tau= \rho_w u_\tau \delta/\mu_w$ is the friction Reynolds number. 
    $Re_\tau^* = \rho_c u_\tau\delta/\mu_c$ is the semi-local Reynolds number. $B_q={q_w}/{\rho_w c_{pw} u_\tau T_w}$ is the heat transfer rate. 
    $Br = {q_w}/{\tau_w U_w}$ is the Brinkman number, which measures the ratio of heat transfer and aerodynamic heating.
    Here, a positive Brinkman number corresponds to heat transfer from the fluid to the wall, and a negative Brinkman number corresponds to heat transfer from the wall to the fluid.}    \begin{tabular}{|c|>{\centering\arraybackslash}m{0.10\textwidth}|
    >{\centering\arraybackslash}m{0.10\textwidth}|
    >{\centering\arraybackslash}m{0.10\textwidth}|
    >{\centering\arraybackslash}m{0.10\textwidth}|
    >{\centering\arraybackslash}m{0.10\textwidth}|
    >{\centering\arraybackslash}m{0.10\textwidth}|
    }
        \hline
                  & D3-05C & D3-05A & D3-05H & D3-10C & D3-10A & D6-05C \\
    \hline
    $Re_\tau$     & 401.09 & 93.99  & 69.16  & 774.80 & 159.85 & 685.05 \\\hline
    $Re_\tau^*$   & 141.02 & 272.63  & 202.16 & 270.51 & 452.81 & 79.97  \\\hline
    $B_q$         & 0.12   & -0.013  & -0.042  & 0.12   & -0.012  & 0.34   \\\hline
    $Br$          & 1.0   & -0.1   & -0.6  & 1.0  & -0.13  & 1.0     \\
    \hline
    \end{tabular}
\end{table}

\subsection{WMLES}

We use the finite volume code CharLES for our WMLESs.
The code solves the filtered full compressible NS equations with a nominally fourth-order accurate discretization in space and a third-order accurate discretization in time.
The sub-grid stress (SGS) is modeled via the dynamic Vreman model \cite{you2007dynamic}.
Further details of the code could be found in Refs. \cite{khalighi2011unstructured,park2014improved} and the references cited therein.
The code has been extensively used for wall-bounded flow calculations.
Recent applications of this code could be found in Refs. \cite{joo2014large, yang2019drag}.
Flow in the wall layer is modeled via the equilibrium wall model detailed in sections \ref{sect:intro} and \ref{sect:derivation}.
We use three wall models, i.e., the EWM in Ref \cite{kawai2012wall}, the semi-local EWM in Ref \cite{yang2018semi}, and the wall model detailed in section \ref{sect:derivation}, whose results are denoted as WMLES$^+$, WMLES$_{\rm YL18}$, and WMLES$^*$.
The grid is of size $N_x\times N_y\times N_z=120\times 40 \times 40$ and resolves a domain of size $L_x\times L_y\times L_z=6\pi\times 2\times 2\pi (\delta)$.
The wall model/LES matching location is at the third off-wall grid, i.e., $h_{wm}/\delta=0.125$, or the first off-wall grid, i.e., $h_{wm}/\delta=0.025$.
While not shown here, we get the same mean flow by doubling the grid numbers in all three directions while keeping $h_{wm}/\delta=0.05$.
This suggests that the errors due to the sub-grid scales are negligible \cite{kawai2012wall,xu2021assessing}.
The setups of the WMLESs are otherwise the same as the DNSs.

\section{Results}
\label{sect:results}

We compare the DNSs, the EWM, where we solve the wall model equations directly, and WMLESs, where we employ the EWM for near-wall turbulence modeling in LES.
EWM$^+$ denotes the conventional viscous units scaled wall model, and EWM$^*$ denotes the semi-local units scaled wall model.
We focus on the temperature statistics. 
A detailed discussion of the velocity results falls out of the scope of this work.

\subsection{Wall model results}

\begin{figure*}
\centering
\includegraphics[width=0.32\textwidth]{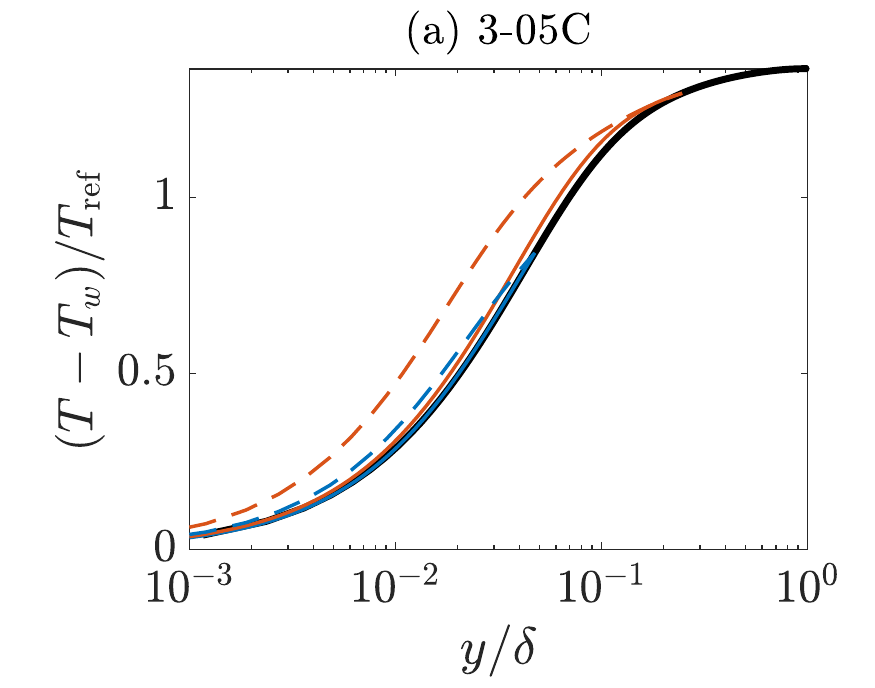}\includegraphics[width=0.32\textwidth]{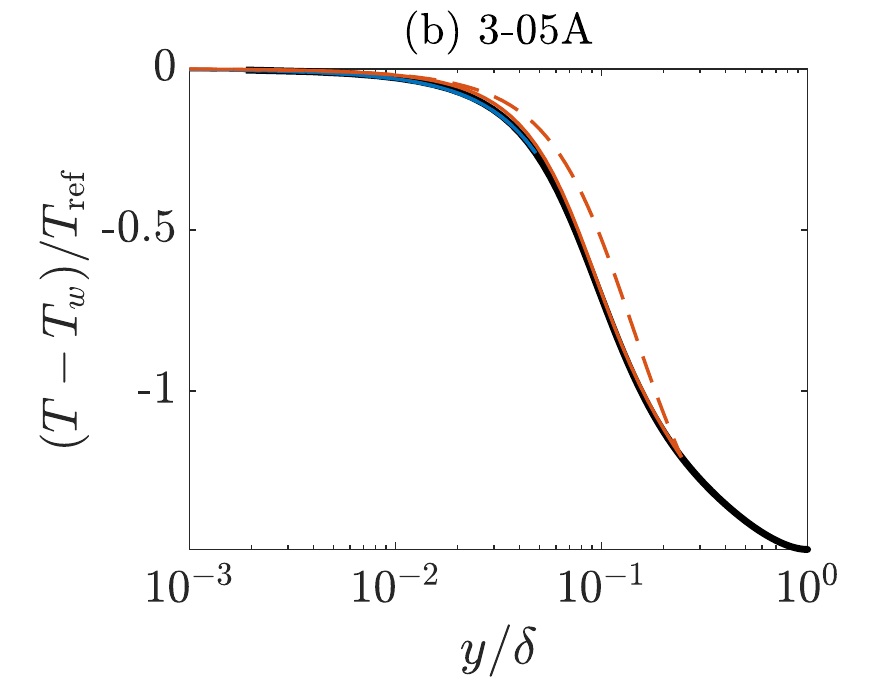}\includegraphics[width=0.32\textwidth]{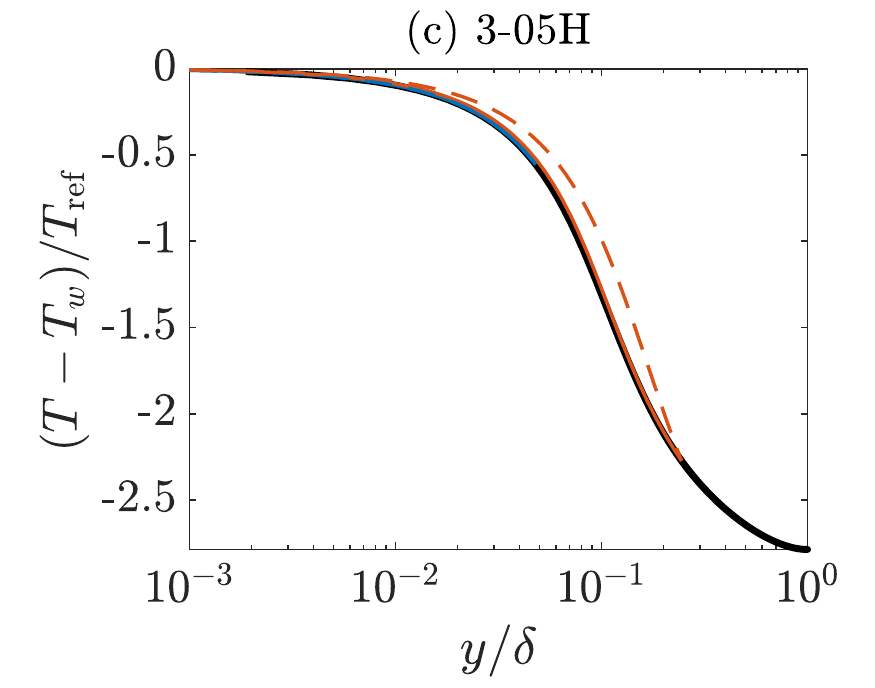}
\includegraphics[width=0.32\textwidth]{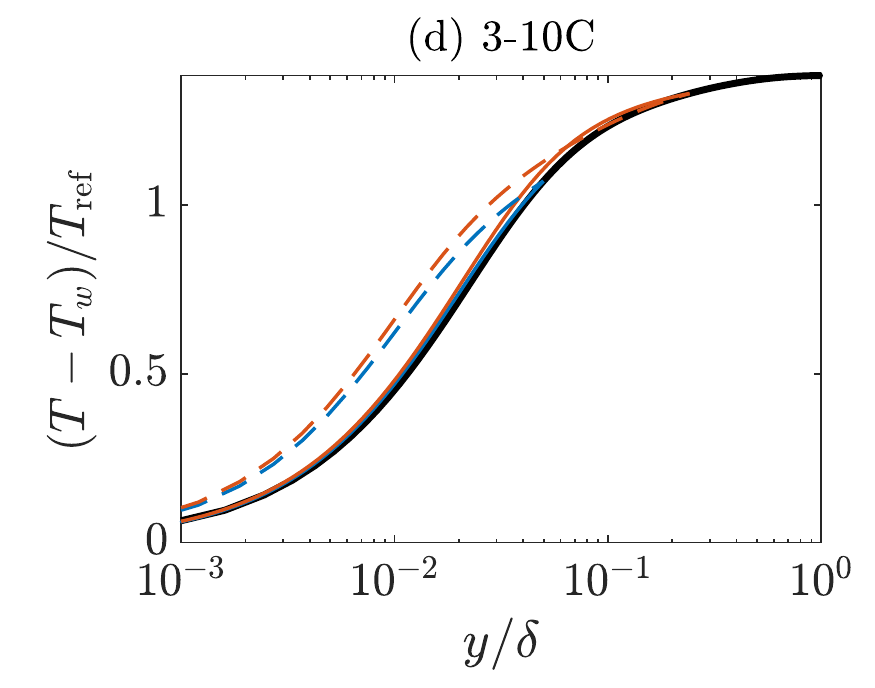}\includegraphics[width=0.32\textwidth]{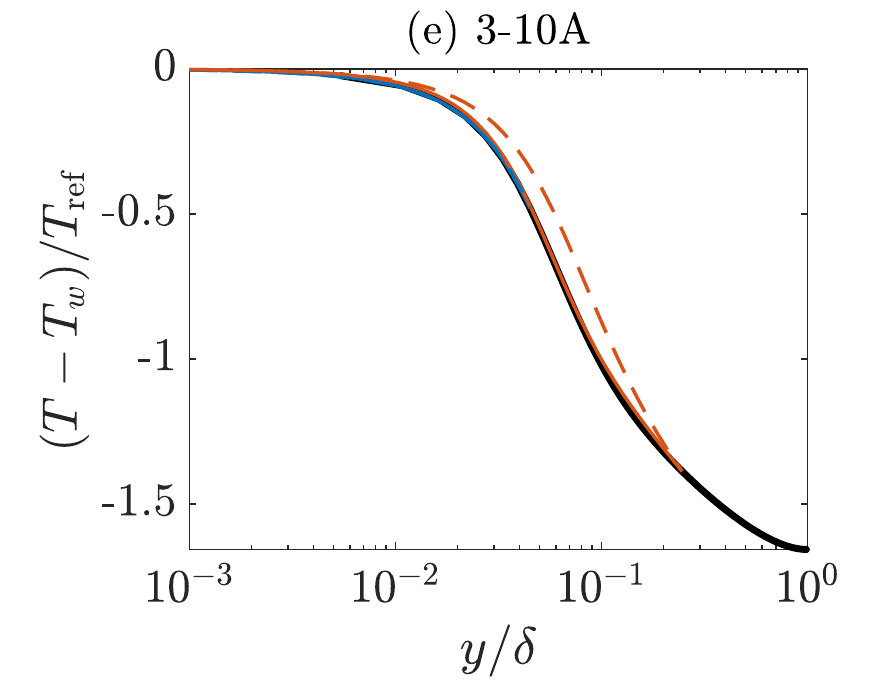}\includegraphics[width=0.32\textwidth]{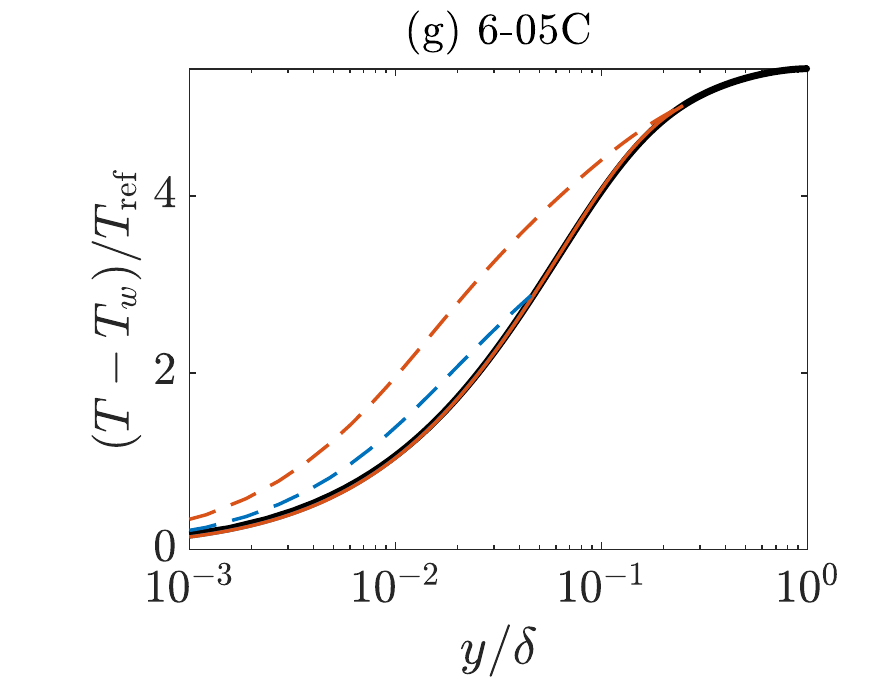}
\includegraphics[width=0.2\textwidth]{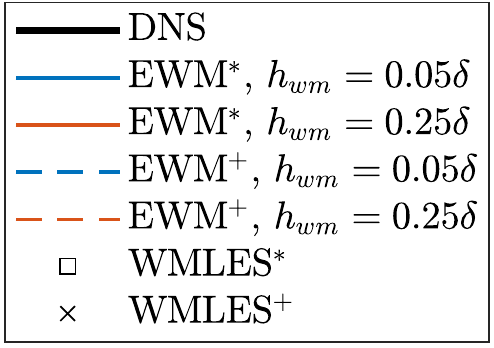}
\caption{Wall model results. 
The temperature profiles are symmetric with respect to the channel centerline, and we show only the results in the lower half channel.
}
\label{fig:T-comp}
\end{figure*}

We solve the wall model equations from the wall, i.e., $y/\delta=0$ up to a distance $y/\delta=0.05$ and 0.25 and compare the results to our DNSs.
This is an {\it a priori} test of our wall model.
Figure \ref{fig:T-comp} compare the results to the DNSs.
EWM$^*$ agrees fairly well with the DNSs for both $h_{wm}/\delta=0.05$ and $0.25$.
Large errors are found in EWM$^+$ when $h_{wm}/\delta=0.25$.
Except for  6-05C and 3-10C, EWM$^+$ agree well with the DNSs when $h_{wm}/\delta=0.05$.

\begin{figure*}
\centering
\includegraphics[width=0.32\textwidth]{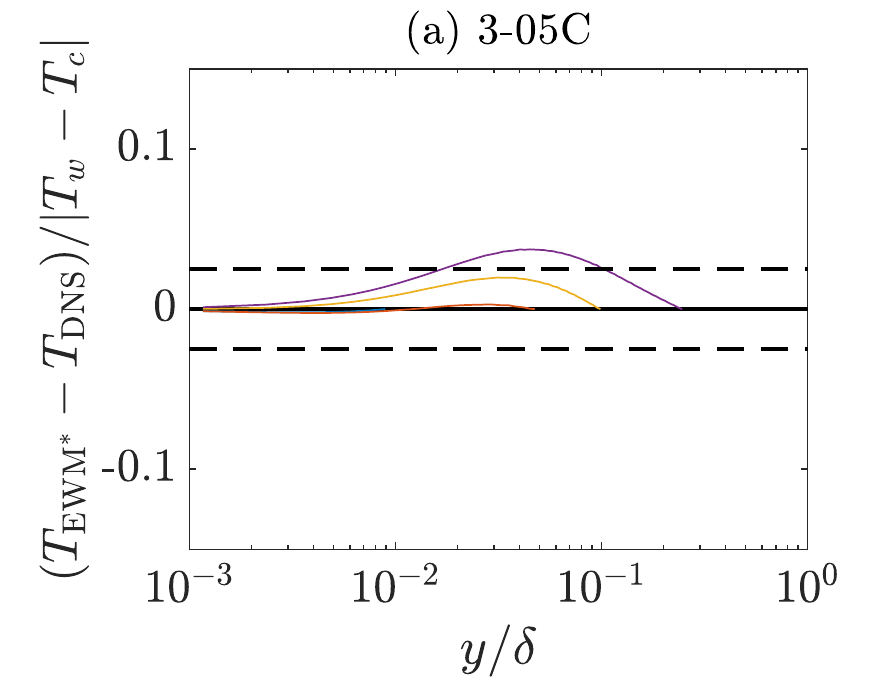}\includegraphics[width=0.32\textwidth]{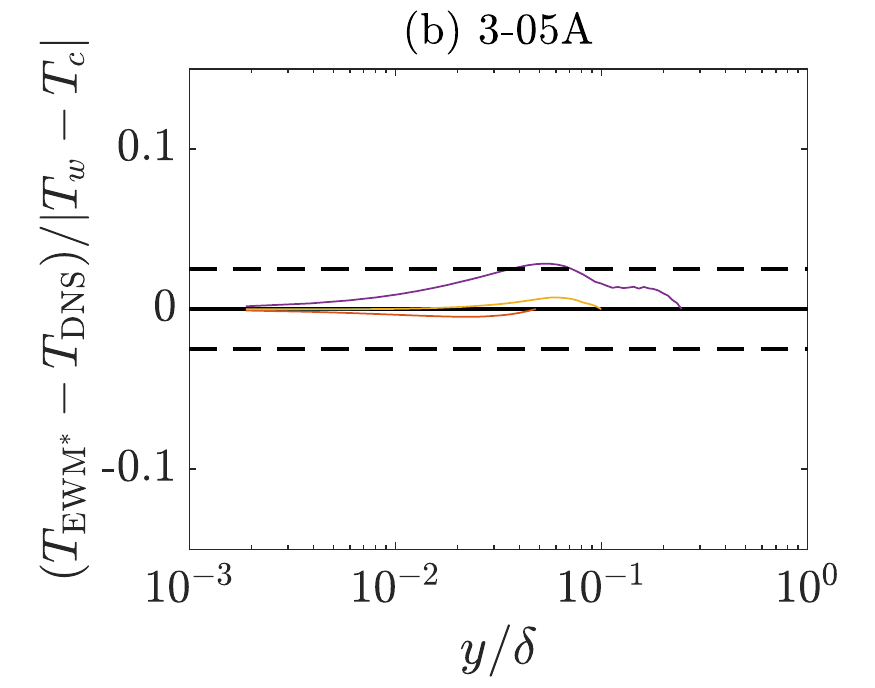}\includegraphics[width=0.32\textwidth]{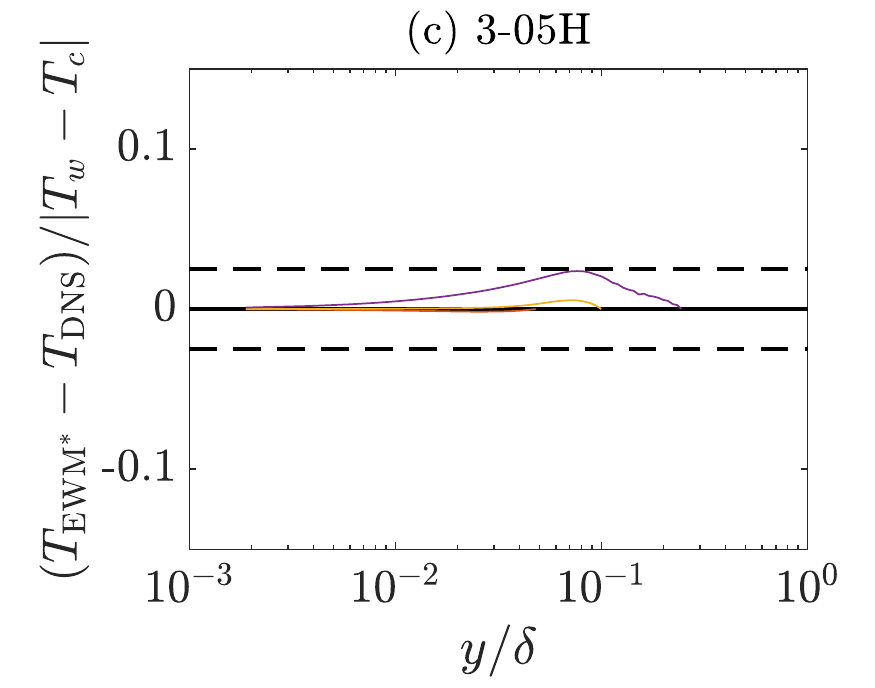}
\includegraphics[width=0.32\textwidth]{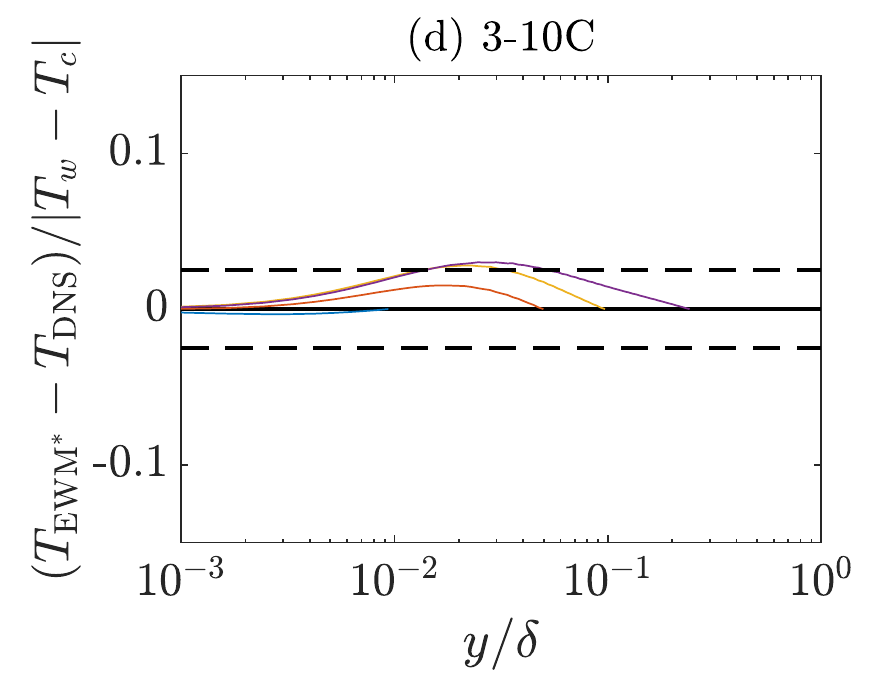}\includegraphics[width=0.32\textwidth]{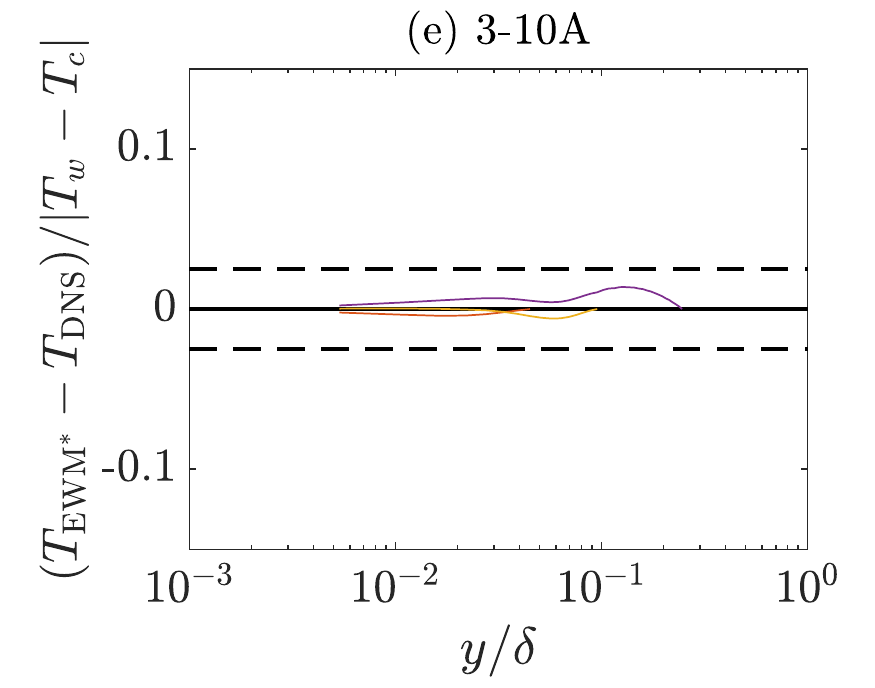}\includegraphics[width=0.32\textwidth]{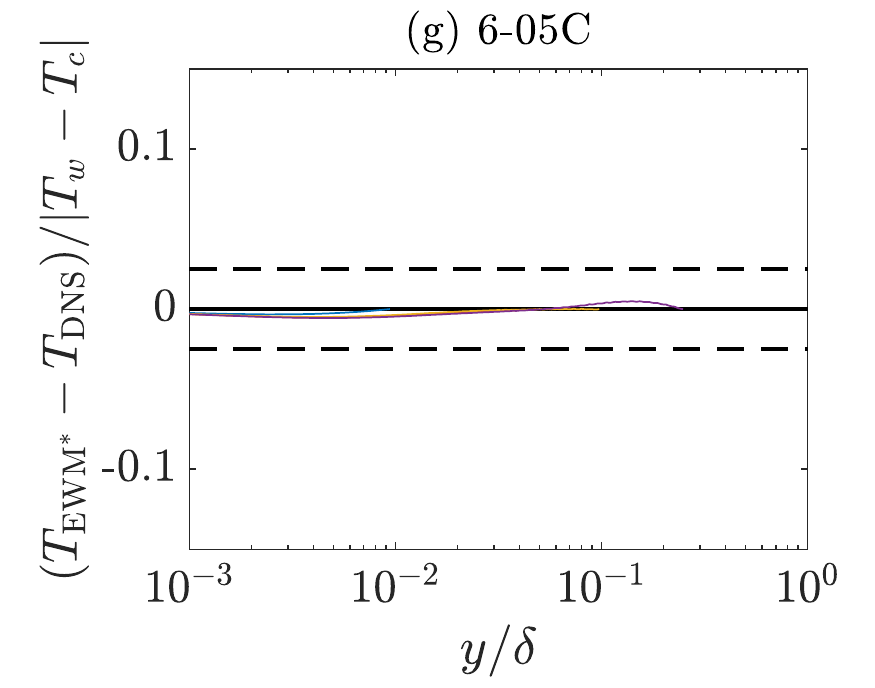}
\includegraphics[width=0.2\textwidth]{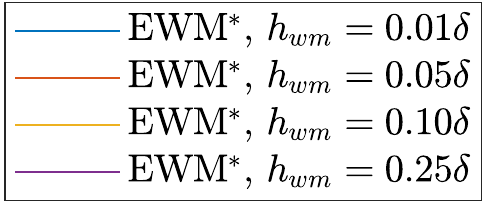}
\caption{Error in EWM$^*$ for all DNS conditions and for $h_{wm}=0.01\delta$, 0.05$\delta$, 0.10$\delta$, and $0.25\delta$. 
The two dashed lines are at $\pm 0.025$.
}
\label{fig:err1}
\end{figure*}

Figure \ref{fig:err1} shows EWM$^*$'s errors for $h_{wm}\delta=0.01$, 0.05, 0.10, and 0.25.
Before we discuss how EWM$^*$ does, we answer the following question: what is an acceptable error in a wall model.
To answer this question, we consider the error due to the von Karman constant $\kappa$.
A conservative estimate of $\kappa$'s uncertainty is 2.5\%, i.e., $\kappa=0.4\pm 0.01$ \cite{nagib2008variations}.
This 2.5\% uncertainty in $\kappa$'s value translates to about a 5\% uncertainty in the wall flux \cite{yang2017log}.
If the von Karman constant is not a significant source of error, a $\pm 2.5\%$ error should be considered acceptable.
Now, we examine the results in figure \ref{fig:err1}.
The error at a given wall-normal height is generally an increasing function of $h_{wm}$, but limiting $h_{wm}<0.25\delta$, the error is, by and large, within $\pm 2.5\%$.
Hence, the results here suggest that, aside from the log layer mismatch, $h_{wm}$ should not be a leading source of error in channel flow calculations.

\subsection{WMLES results}

We compare WMLESs directly to DNSs.
This is an {\it a posteriori} test.
The discussion in this section is limited to the channel flow configuration.
While boundary-layer flow statistics at one (a few) streamwise locations in supersonic boundary layers is publicly available \cite{zhang2018direct,volpiani2018effects}, a fair comparison between WMLESs and these DNSs is highly non-trivial because of a lack of knowledge of boundary layer's Reynolds number scalings at high speeds as well as a lack of knowledge of the inflow condition.
The above difficulty has also motivated the H and A cases in table \ref{tab:DNS}, where we include a heat source in a channel.

\begin{figure*}
\centering
\includegraphics[width=0.32\textwidth]{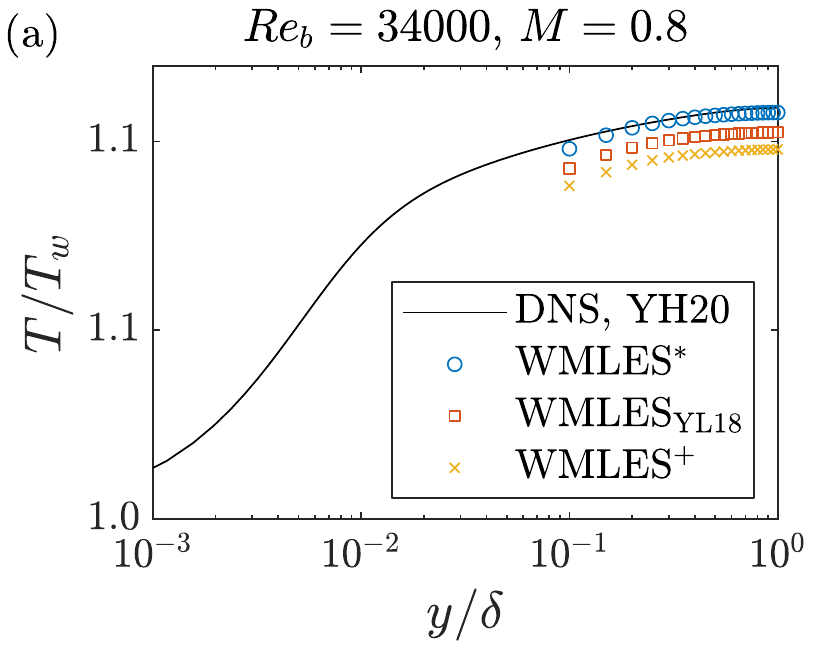}~~\includegraphics[width=0.32\textwidth]{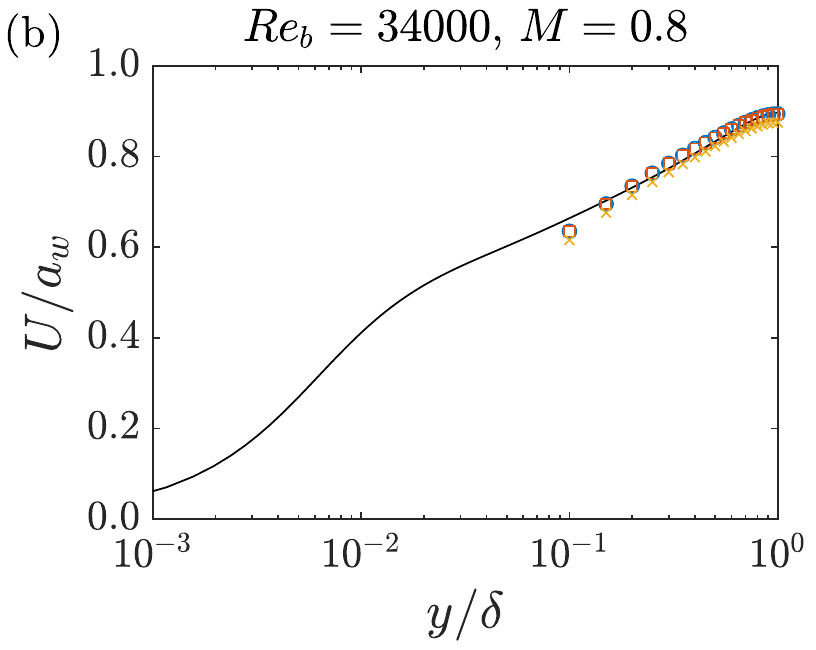}
\includegraphics[width=0.32\textwidth]{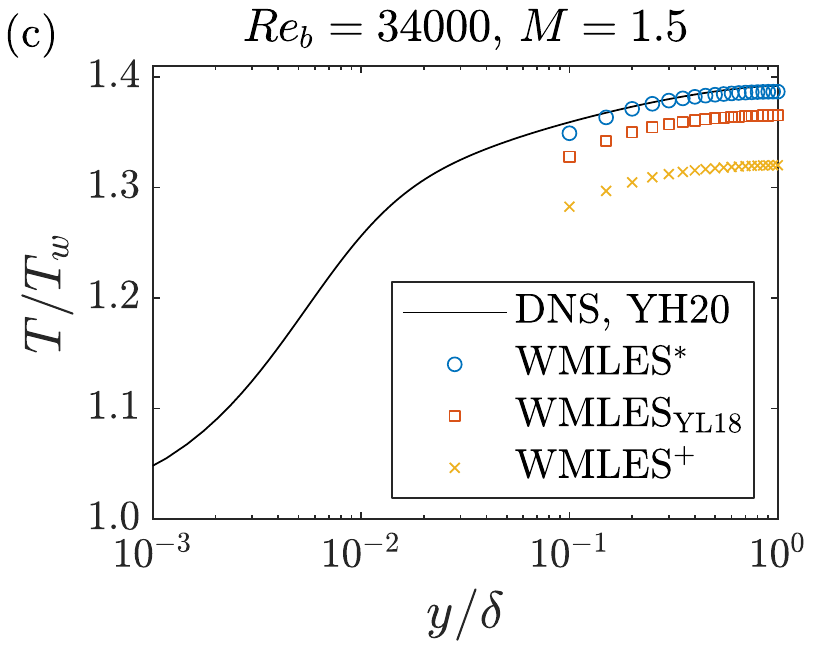}~~\includegraphics[width=0.32\textwidth]{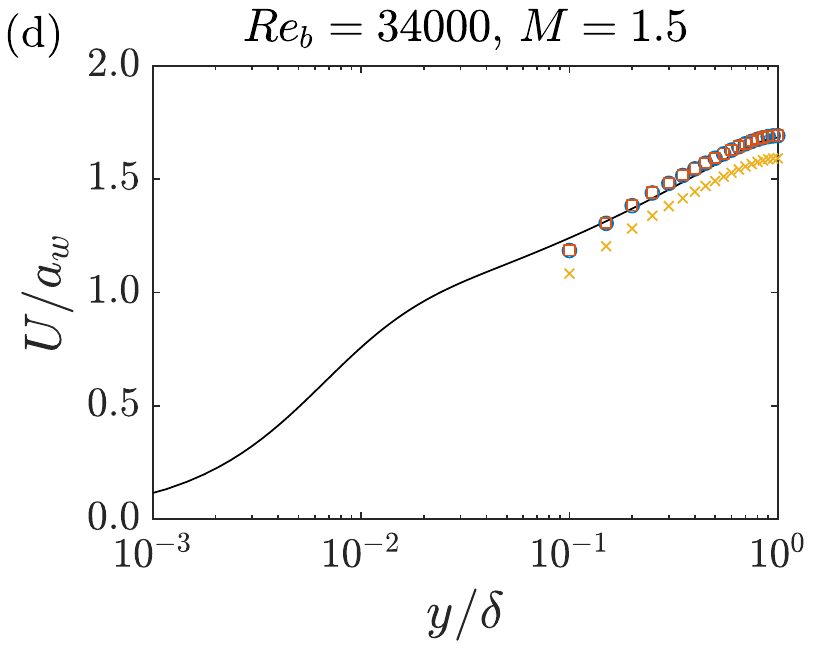}
\caption{(a) $T/T_w$ as a function of $y/\delta$ for the channel flow DNS in Ref \cite{yao_supersonic_2019} and for WMLES$^*$, WMLES$_{\rm YL18}$, and WMLES$^+$.
The flow condition is $Re_b=34000$, $M=0.8$.
(b) Same as (a) but for $U/a_w$.
(c) Same as (a) but at the flow condition $Re_b=34000$, $M=1.5$.
(d) Same as (c) but for $U/a_w$.
}
\label{fig:WLhiRe}
\end{figure*}

\begin{figure*}
\centering
\includegraphics[width=0.32\textwidth]{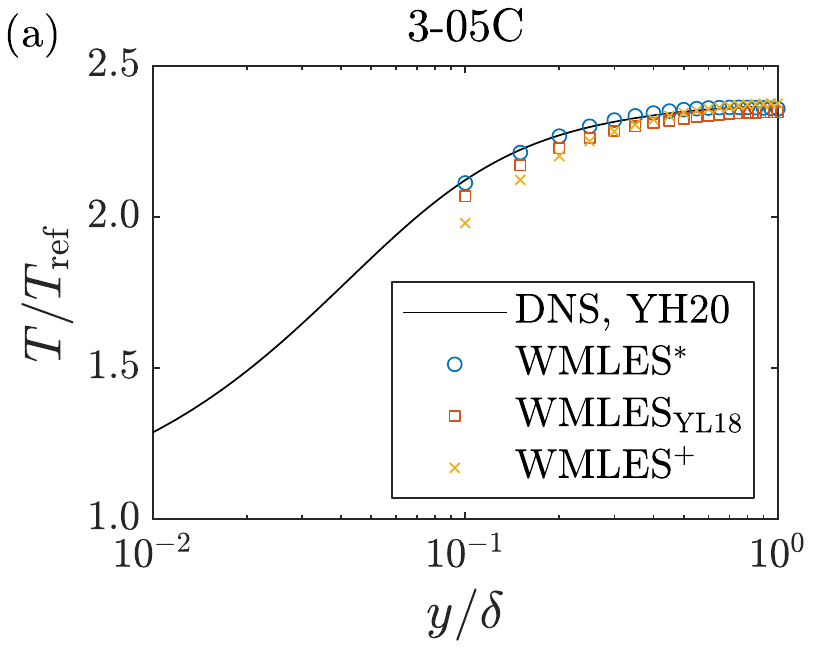}~~\includegraphics[width=0.32\textwidth]{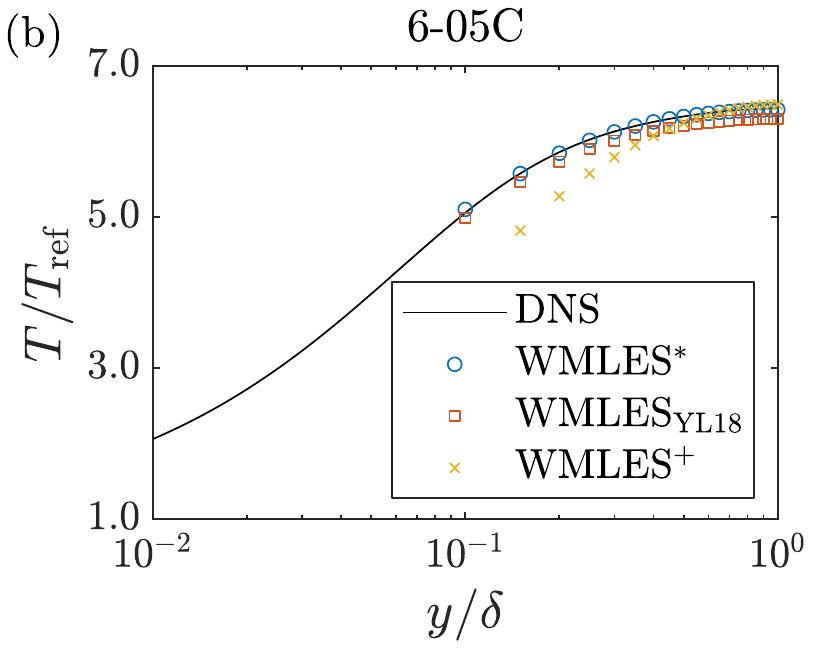}~~\includegraphics[width=0.32\textwidth]{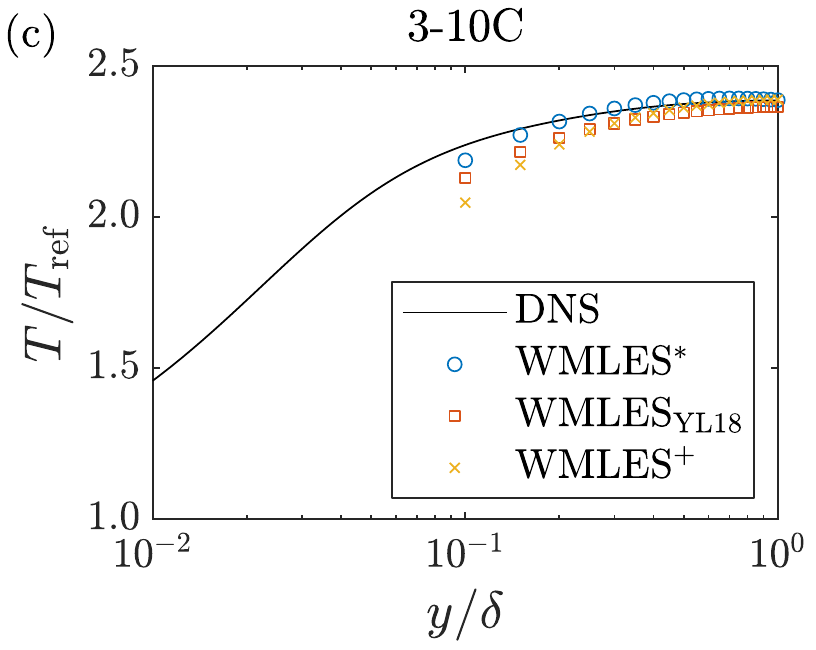}
\caption{$T/T_{\rm ref}$ as a function of $y/\delta$ at the flow condition (a) 3-05C, (b) 6-05C, (c) 3-10C.
}
\label{fig:WLC}
\end{figure*}

Figure \ref{fig:WLhiRe} compares WMLESs and the two channel flow DNSs in Ref \cite{yao_supersonic_2019}.
The two DNSs are at a bulk Reynolds number $Re_b=34000$, and the bulk Mach numbers are $M=0.8$, 1.5.
For DNSs of supersonic channel flow, $Re_b=34000$ is a high Reynolds number.
Figure \ref{fig:WLhiRe} compares both $U$ and $T$.
We make the following observations.
First and foremost, we see a logarithmic layer in (b, d).
Examining the velocity $U/a_w$, the conventional viscous scaled wall model, i.e., WMLES$^+$ gives an accurate estimate of the velocity at the subsonic speed M$=0.8$ but underestimates the velocity at the supersonic speed M$=1.5$; the model in Ref. \cite{yang2018semi}, i.e., WMLES$_{\rm YL18}$, gives an accurate estimate of the velocity at both M$=0.8$ and 1.5.
This is consistent with Ref. \cite{yang2018semi}.
The present model, i.e., WMLES$^*$, does not seem to have an effect on the velocity.
Examining the temperature $T/T_w$, WMLES$_{\rm YL18}$ follows the DNSs more closely than WMLES$^+$, WMLES$^*$ more closely than WMLES$_{\rm YL18}$, and WMLES$^*$ agree well with the DNSs at both the subsonic speed M$=0.8$ and the supersonic speed M$=1.5$.
The errors in $T-T_w$ are 10\% and 6\% in WMLES$^+$ and WMLES$_{\rm YL18}$ at the channel centerline for the M$=0.8$ case.
For the M$=1.5$ case, the errors are 18\% and 7\% .
These results validate our model.
Meanwhile, they explain why Yang and Lv were able to get accurate velocity estimates without Eq. \eqref{eq:Prt-new} \cite{yang2018semi}.
Figure \ref{fig:WLC} compares WMLESs and DNSs for the C cases in table \ref{tab:DNS}.
For brevity, we show only the thermal field results.
The results are similar to these in figure \ref{fig:WLhiRe}---although the difference between different wall models is less apparent.
While not shown here, the first grid point implementation \cite{yang2017log} and the third grid point implementation \cite{kawai2012wall} lead to the same results.

\begin{figure*}
\centering
\includegraphics[width=0.32\textwidth]{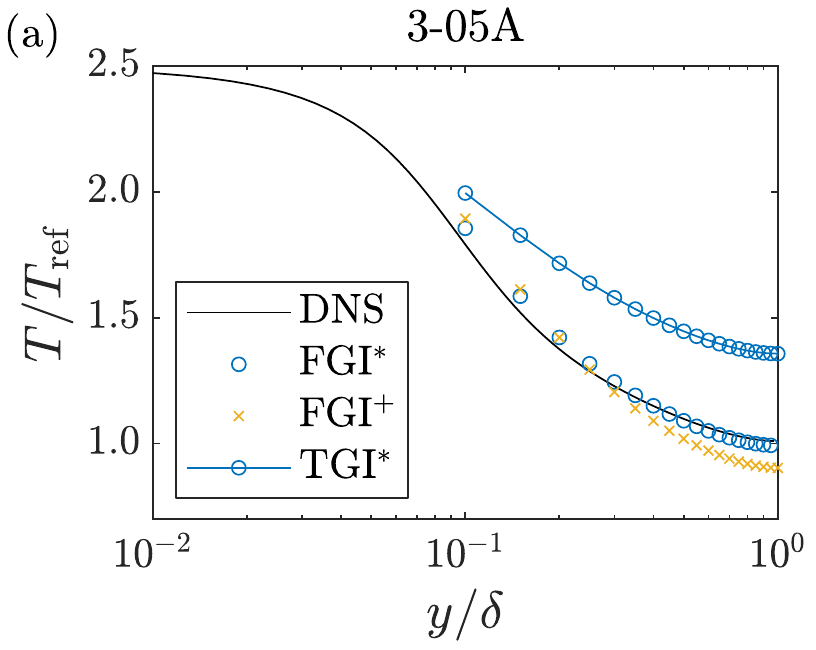}~~\includegraphics[width=0.32\textwidth]{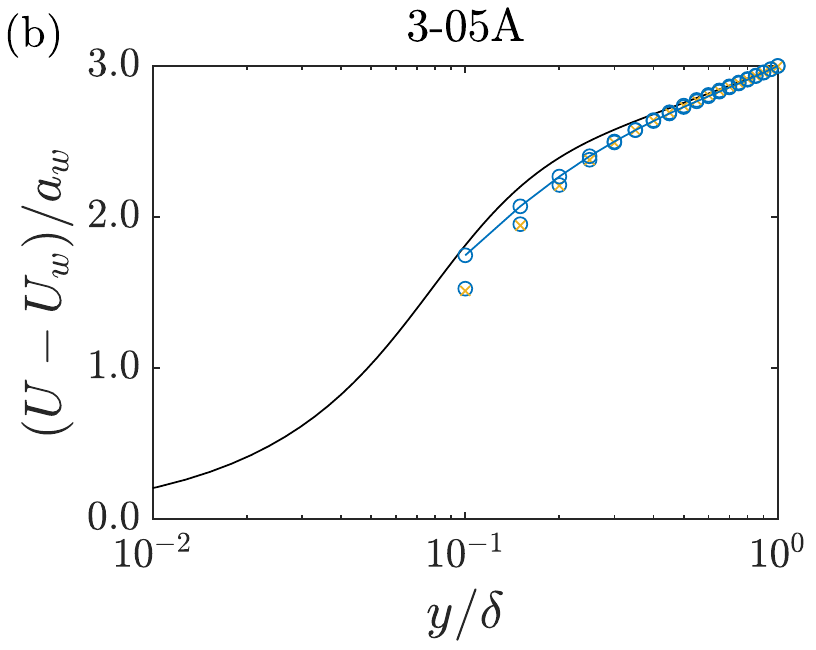}
\includegraphics[width=0.32\textwidth]{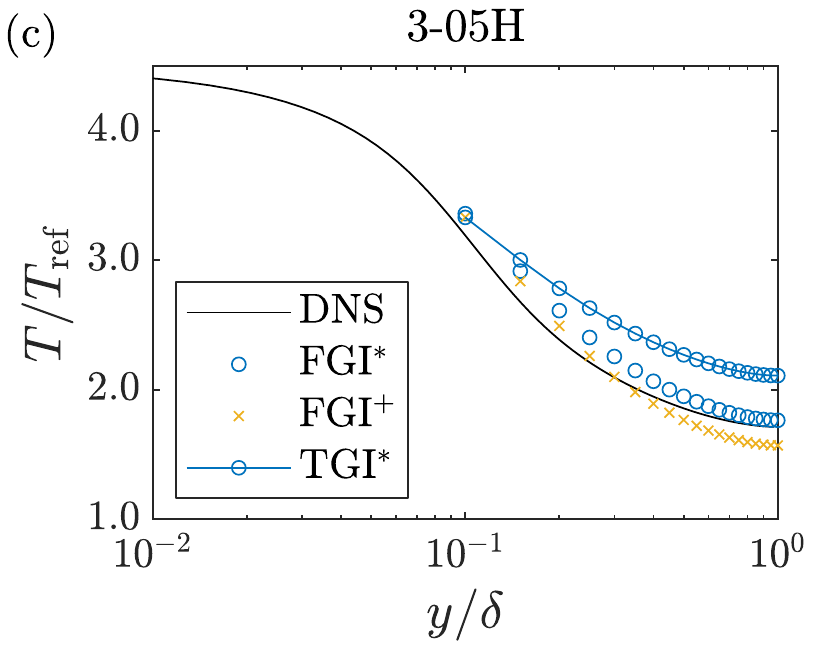}~~\includegraphics[width=0.32\textwidth]{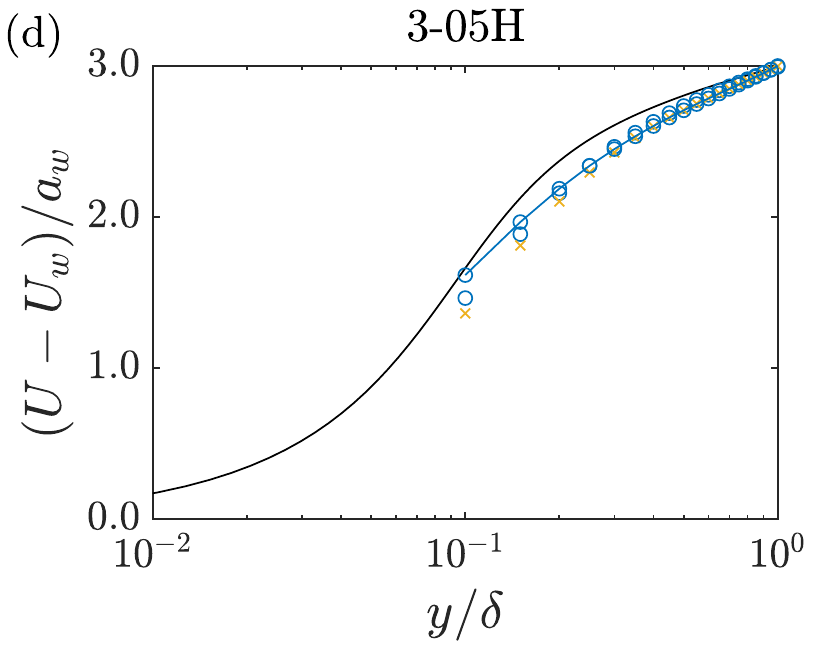}
\caption{$T/T_{\rm ref}$ as a function of $y/\delta$ at the flow condition (a) 3-05A, (c) 3-05H.
$(U-U_w)/a_w$ as a function of $y/\delta$ at the flow condition (b) 3-05A, (d) 3-05H.
}
\label{fig:WLHA}
\end{figure*}

Next, we compare WMLESs and DNSs for the A (nearly adiabatic wall) and the H (hot wall) cases.
WMLES$^+$, WMLES$_{\rm YL18}$, and WMLES$^*$ differ in a similar way as in the C cases, and we do not repeat the discussion for brevity.
Interestingly, the first grid point implementation and the third grid point implementation lead to different results for the A and H cases.
Figure \ref{fig:WLHA} shows the results. 
We show WMLES$^*$ with the first grid point implementation, i.e., FGI$^*$, and WMLES$^*$ with the third grid point implementation, i.e., TGI$^*$. 
WMLES$^+$ with the first grid point implementation, i.e., FGI$^+$, is included for comparison.
Examining the velocity $(U-U_w)/a_w$ in (b, d), FGI$^*$ and TGI$^*$ give similar velocity estimates, and they both agree reasonably well with the DNSs.
Examining the temperature $T/T_{\rm ref}$ in (a, c), FGI$^*$ gives reasonably accurate estimates of $T$ at both wall conditions, but TGI$^*$ significantly overestimates $T$.
This directly contradicts the {\it a priori} test results in figure \ref{fig:err1}---although the focuses of the {\it a priori} test and the {\it a posteriori} test are different.

While there can be other factors, the following factors are responsible for the difference between FGI$^*$ and TGI$^*$.
First, FGI places the wall-model/LES matching location closer to the wall than TGI; as a result, FGI practically ``sees'' a higher near-wall resolution than TGI.
Considering first that the semi-local scaling works not so well for adiabatic and hot walls, and second that $dy^*/dy$ is an increasing function of $y$, safely bringing the wall-model/LES matching location close to the wall is undoubtedly beneficial.
This is like WRLES and WMLES: increased near-wall resolution benefits.
In addition, not accounting for the heat source is responsible too.
The wall heat flux is $q_w=U_w\tau_w+\phi \delta$.
FGI and TGI predict similar $\tau_w$ and similar $q_w$.
Rewriting the energy balance between the wall and the matching location, the wall heat flux is $q_w=F_{h_{wm}}+U_w\tau_w+\phi h_{wm}$, where $F_{h_{wm}}$ is the heat flux at $h_{wm}$.
Because the effect of the source term increases as $h_{wm}$ increases, the effect of not accounting for it in the wall model increases as $h_{wm}$ increases, which is what we see in figure \ref{fig:WLHA}.
In spite of the above consideration, incorporating source terms in the wall model equations is not a good idea for the following two reasons.
First, the wall model equations rely on the constant stress layer assumption, and under that assumption, any source term should be negligible.
The second and the more important reason is that, if one were to include a source term, one would have to recalibrate all other model coefficients, which is likely to be a significant source of uncertainty.
In fact, including the mean pressure gradient in the momentum equation, the present wall model would not even be able to accurately predict the log law.

Further details of the WMLESs, including the wall heat flux and the wall heat transfer rate, are reported in the appendix.
In all, we see that the eddy conductivity in Eq. \eqref{eq:semi-alphat} works very well in the context of WMLES.
This is in direct contrast to the results in figure \ref{fig:semi-comp}, where the semi-local scaling fails catastrophically.

\section{Discussion}
\label{sect:discussion}

In this section, we answer the following question:
why the semi-local scaling fails in collapsing the high-speed data but the resulting wall model works well.
We will show that some of the assumptions in Ref \cite{patel2017scalar} do not necessarily apply to high-speed flows.

\begin{figure*}
\centering
\includegraphics[width=0.85\textwidth]{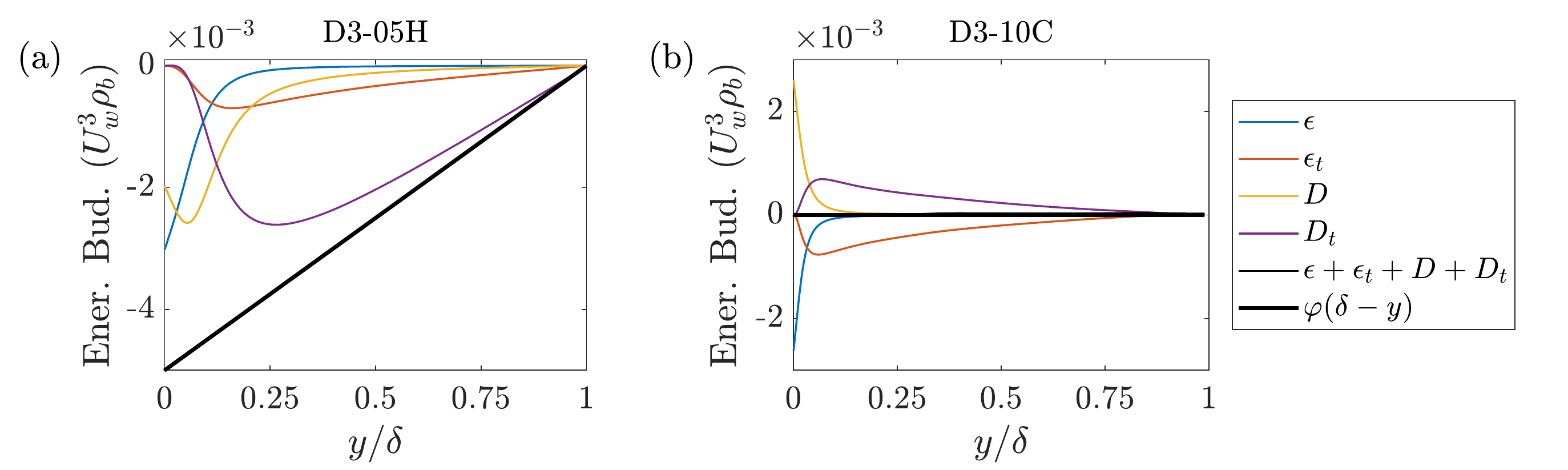}
\caption{Energy budget terms in (a) D3-05H, where the heat source term is non-zero, and (b) D3-10C, where the heat source term is $\varphi=0$. 
}
\label{fig:ener-bud}
\end{figure*}

First, we examine the Reynolds averaged energy equation:
\begin{linenomath*}\begin{equation}
  \dfrac{d}{d y}\left(\overline{ \sigma_{yj} u_j} -\overline{\rho v \frac{1}{2}u_ju_j } -\overline{\rho v c_p T}+\overline{\lambda \frac{\partial T }{\partial y}}\right) + \varphi = 0,
  \label{eq:energy}
\end{equation}\end{linenomath*}
where $u$, $v$, $w$ are the instantaneous velocity in the three Cartesian directions, $\sigma_{ij}$ is the instantaneous viscous stress tensor.
The reader is directed to Ref \cite{morinishi2004direct} for more details of the energy equation.
Integrating Eq. \eqref{eq:energy} in $y$ from the channel centerline to a distance $y$ from the wall, we get
\begin{linenomath*}\begin{equation}
\begin{split}
    \epsilon+\epsilon_t+D+D_t=(\delta-y)\varphi,
    \label{eq:energy2}
\end{split}
\end{equation}\end{linenomath*}
where $\delta \varphi=q_w-U_w\tau_w$.
Here, $\epsilon=\overline{ \sigma_{y j} u_j}$, $\epsilon_t$ $=- \overline{\rho v \frac{1}{2}u_ju_j}$, $D=\overline{\lambda {\partial T }/{\partial y} }$, and $D_t=- c_p\overline{\rho v T}$.
Figure \ref{fig:ener-bud} shows the terms in the energy equation for the cases D3-05H and D3-10C, where the source term $\varphi$ is non-zero and zero in (a, b) respectively.
The sums of the four terms are $(\delta - y)\varphi$ in the case D3-05H and $0$ in the case D3-10C.
We see from figure \ref{fig:ener-bud} that the viscous heating, i.e., $\epsilon$ and $\epsilon_t$, are non-negligible in wall layer for $y/\delta\lesssim 0.2$.

We take a reference frame such that $U_w=0$.
In the constant stress layer, $y\ll \delta$ and Eq. \eqref{eq:energy2} becomes
\begin{linenomath*}\begin{equation}
\begin{split}
    \epsilon +\epsilon_t + D + D_t=\delta\varphi=q_w.
    \label{eq:energy3}
\end{split}
\end{equation}\end{linenomath*}
Comparing Eq. \eqref{eq:energy3} and the energy equation in Ref. \cite{patel2017scalar},
Patel et al. did not account for aerodynamic heating, i.e., $\epsilon$ and $\epsilon_t$.
Now, comparing Eq. \eqref{eq:energy3} and the wall model equation Eq. \eqref{eq:WM-T}, the EWM models the four terms in Eq. \eqref{eq:energy3} as follows
\begin{linenomath*}\begin{equation}
\begin{split}
    \epsilon=\mu\frac{du_{||}}{dy}u_{||}, ~~\epsilon_t=\mu_t\frac{du_{||}}{dy} u_{||},~~
    D=\lambda \frac{dT}{dy}, ~~D_t=\frac{c_p\mu_t}{Pr_t} \frac{dT}{dy}.
\end{split}
\end{equation}\end{linenomath*}
Considering that both $\epsilon$ and $\epsilon_t$ are non-negligible, the fact that the EWM models aerodynamic heating but Patel et al. did not partly explains why the EWM works well but the semi-local scaling fails.

\begin{figure}
\centering
\includegraphics[width=0.32\textwidth]{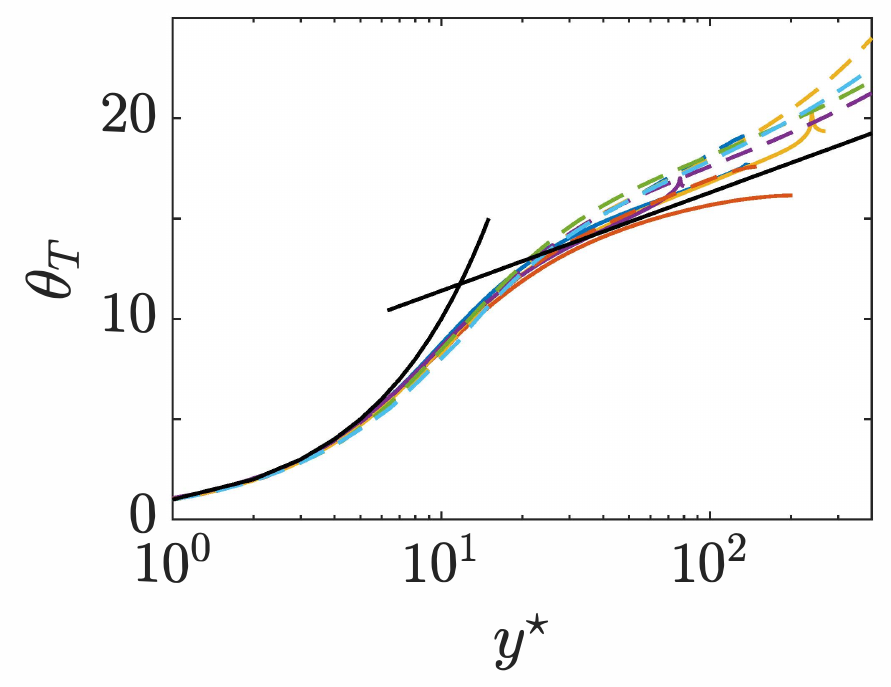}
\caption{Same as figure \ref{fig:T-comp} but for the scaling in Eqs. \eqref{eq:new4} and \eqref{eq:semi-T2}, which accounts for aerodynamic heating.
}
\label{fig:new-comp}
\end{figure}

To further verify our conclusion, we incorporate aerodynamic heating in the semi-local scaling and examine if the resulting scaling does a better job collapsing the high-speed data.
We follow Ref \cite{patel2017scalar} and re-write Eq. \eqref{eq:energy3}:
\begin{linenomath*}\begin{equation}
    \frac{\epsilon+\epsilon_t}{q_w}+\left(\frac{\alpha_t}{\mu}+\frac{1}{Pr^*}\right)\frac{\delta}{Re^*_\tau}\sqrt{\frac{\rho}{\rho_w}}\frac{d\theta^+}{dy}=1
    \label{eq:new1}
\end{equation}\end{linenomath*}
Rearranging the above equation, we have 
\begin{linenomath*}\begin{equation}
\small
\frac{h}{Re^*_\tau}\sqrt{\frac{\rho}{\rho_w}}\frac{dy^*}{dy}\frac{d\theta^+}{dy^*}\big/\left(1-\frac{\epsilon+\epsilon_t}{q_w}\right)=1\big/\left(\frac{\alpha_t}{\mu}+\frac{1}{Pr^*}\right).
\label{eq:new2}
\end{equation}\end{linenomath*}
Patel et al. argue that the right-hand side of Eq. \eqref{eq:new2} is a function of $y^*$ when given the molecular Prandtl number.
This is the universality assumption, and we retain this assumption.
It follows that we may define $\theta^*$ such that
\begin{linenomath*}\begin{equation}
\frac{d\theta^*}{dy^*}=1\big/\left(\frac{\alpha_t}{\mu}+\frac{1}{Pr^*}\right).
\label{eq:new3}
\end{equation}\end{linenomath*}
Equations \eqref{eq:new2} and \eqref{eq:new3} give
\begin{linenomath*}\begin{equation}
\begin{split}
    \frac{d\theta^*}{d\theta^+}=\frac{d\theta^*}{dy^*}\frac{dy^*}{d\theta^+}=\frac{\delta}{Re^*_\tau}\sqrt{\frac{\rho}{\rho_w}}\frac{dy^*}{dy}\big/\left(1-\frac{\epsilon+\epsilon_t}{q_w}\right)
    =\left(1+\frac{y^*}{Re_\tau^*}\frac{dRe_\tau^*}{dy}\right)\sqrt{\frac{\rho}{\rho_w}}\big/\left(1-\frac{\epsilon+\epsilon_t}{q_w}\right),
\end{split}
\label{eq:new4}
\end{equation}\end{linenomath*}
Equations \eqref{eq:new4} and \eqref{eq:semi-T2} together define a new semi-local scaling.
The new scaling contains a new term $1-(\epsilon+\epsilon_t)/q_w$ that accounts for aerodynamic heating.
To accurately evaluate these terms, the energy equation must be statistically converged.
This requirement prevents us from utilizing some of the publicly available data.
We apply this new semi-local scaling to the data in figure \ref{fig:semi-comp}.
Figure \ref{fig:new-comp} shows the results.
Comparing figures \ref{fig:semi-comp} and figure \ref{fig:new-comp}, the new semi-local scaling collapses the high speed data considerably better than the one in Ref \cite{patel2017scalar}.
Hence, we conclude that the semi-local scaling in Ref. \cite{patel2017scalar} fails at collapsing the high-speed data because it does not account for aerodynamic heating.
(Again, this is not a criticism. In Ref \cite{patel2017scalar}, the flow is at low speeds, and therefore aerodynamic heating is indeed negligible.)

Having answered why the semi-local scaling in Ref \cite{patel2017scalar} fails, we are yet to answer why our wall model works.
In the following, we show that the new semi-local scaling that incorporates aerodynamic heating leads to the same eddy conductivity as the semi-local scaling in Ref \cite{patel2017scalar}.
We follow the same steps as in section \ref{sect:derivation}:
\begin{linenomath*}\begin{equation}
\begin{split}
    \frac{\alpha_t}{\mu}&=\frac{-\rho \overline{v''\theta''}}{\mu\dfrac{d\theta}{dy}}=\frac{q_w\left(1-\dfrac{\epsilon+\epsilon_t}{q_w}\right)}{\dfrac{d\theta}{d\theta^*}\dfrac{d\theta^*}{d\theta_T}\dfrac{d\theta_T}{dy^*}\dfrac{dy^*}{dy}}-\frac{1}{Pr^*}\\
    &=\frac{q_w}{q_w\dfrac{d\theta_T}{dy^*}\dfrac{1+\alpha_t/\mu}{1/Pr^*+\alpha_t/\mu}\dfrac{1}{\mu}}-\frac{1}{Pr^*}\\
    &=\frac{\left( 1+\kappa_T y^*D_T\right)\left({\mu}/{Pr^*} + \alpha_t\right)}{\mu+{\alpha_t}} -\frac{1}{Pr^*}.
\end{split}
\label{eq:Prt-new}
\end{equation}\end{linenomath*}
Equation \eqref{eq:Prt-new} is again a quadratic equation of $\alpha_t/\mu$.
Its solution is just Eq. \eqref{eq:semi-alphat}.
Here, $\rho \overline{v''\theta''}=q_w-\epsilon-\epsilon_t-D$ according to Eq. \eqref{eq:energy3}, and $d\theta^*/d\theta^+$ is given by Eq. \eqref{eq:new4}.
In conclusion, incorporating aerodynamic heating or not, the semi-local scaling gives the same eddy conductivity, and this is why the wall model had worked.

\section{Conclusions}
\label{sect:conclusions}

We exploit the semi-local scaling and derive an eddy conductivity closure for the EWM.
The wall distance scaling in the damping functions depends on the universal laws of the wall, in this case, the semi-local scalings.
The derivation leads to the following turbulent Prandtl number in the sublayer and the logarithmic layer
\begin{linenomath*}\begin{equation*}
    Pr_t=\frac{\kappa}{\kappa_T}\left[\frac{1-\exp(-y^*/A)}{1-\exp(-y^*/A_T)}\right]^2,
\end{equation*}\end{linenomath*}
where $y^*=\sqrt{\rho \tau_w} y/\mu$ is the semi-local scaled distance from the wall.
Comparing WMLESs to DNSs, we conclude: first, the thermal field is more challenging to model and compute than the momentum field; second, {\it a priori} tests and {\it a posteriori} tests do not necessarily agree with each other; third, the present model gives similar velocity predictions as the model in Ref \cite{yang2018semi}; fourth, the present wall model gives more accurate results than both the model in Ref \cite{yang2018semi} and the conventional viscous scaled model; fifth, the first grid point implementation in Ref \cite{yang2017log} gives more accurate results than the $n$th ($n>1$) grid point implementation in Ref \cite{kawai2012wall} for adiabatic and hot walls.

The success of semi-local scaling in the WMLES and the poor collapse of the high-speed DNS data begs the question.
Detailed analysis of the energy budget shows that the EWM works because it models, albeit very crudely, aerodynamic heating, a term that  is missing in the semi-local scaling \cite{patel2017scalar}.
By accounting for aerodynamic heating in semi-local scaling, we show that the new semi-local scaling collapses the DNS data considerably better.
We also show that the new semi-local scaling gives rise to exactly the same eddy conductivity as the semi-local scaling in Ref \cite{patel2017scalar}, thereby answering the above question. 




\section*{Acknowledgement}
The DNSs are conducted on Tianhe \uppercase\expandafter{\romannumeral2} at Guangzhou.
The WMLESs are conducted on  ACI-ICS at Penn State. 
CP acknowledges Xinliang Li for providing the DNS code. 
CP acknowledges Zhou Jiang for fruitful discussion.
XY acknowledges Penn State for financial support.
Shi acknowledges financial support from NSFC 91752202.

\clearpage
\newpage

\begin{appendices}
\section{Further WMLES details}
\label{app:WMLES}

\begin{table}
    \caption{\label{tab:results}Wall shear stresses $\tau_w\times 10^2/(\rho_b U_w^2)$ and the wall heat fluxes $q_w\times 10^2/\rho_b U_w^3$.
A positive heat flux is a flux from the fluid to the wall, and a negative heat flux is a flux from the wall to the fluid.
For all the numbers reported in the table, we keep two digits behind the decimal point.}
    \begin{tabular}{|cc|
    >{\centering\arraybackslash}m{0.10\textwidth}|
    >{\centering\arraybackslash}m{0.10\textwidth}|
    >{\centering\arraybackslash}m{0.10\textwidth}|
    >{\centering\arraybackslash}m{0.10\textwidth}|
    >{\centering\arraybackslash}m{0.10\textwidth}|
    }
    \hline
\multicolumn{1}{|l}{}    & \multicolumn{1}{l|}{} & 03-05C             & 03-05A               & 03-05H               & 03-10C                                         & 06-05C             \\ \hline
\multicolumn{2}{|l|}{DNS}                        & \multicolumn{1}{c|}{0.28, 0.28}       & \multicolumn{1}{c|}{0.26, -0.03}      & \multicolumn{1}{c|}{0.31, -0.20 }     & \multicolumn{1}{c|}{0.25, 0.25 }         & \multicolumn{1}{c|}{0.32, 0.32}       \\\hline
\multicolumn{2}{|l|}{WMLES$^*$}                  & \multicolumn{1}{c|}{0.29, 0.29} & \multicolumn{1}{c|}{0.26, -0.04} & \multicolumn{1}{c|}{0.32, -0.19} & \multicolumn{1}{c|}{0.25, 0.25} &  \multicolumn{1}{c|}{0.33, 0.33} \\\hline
\multicolumn{2}{|l|}{WMLES$^+$}                  & \multicolumn{1}{c|}{0.42, 0.42} & \multicolumn{1}{c|}{0.25, -0.04} & \multicolumn{1}{c|}{0.29, -0.21} & \multicolumn{1}{c|}{0.35, 0.35} &  \multicolumn{1}{c|}{0.51, 0.51} \\ \hline
    \end{tabular}
\end{table}

Table \ref{tab:results} shows the wall heat flux $q_w/\rho_b U_w^3$ and $\tau_w/\rho_b U_w^2$ for DNSs, and WMLESs.
Here, $\tau_w/\rho_b U_w^2$ is the skin friction coefficient $C_f$; we report the heat flux $q_w/\rho_b U_w^3$ instead of the heat transfer rate $B_q=q_w/\tau_w U_w$ because the errors in $\tau_w$ and $q_w$ may cancel.
\end{appendices}

\clearpage
\newpage

\bibliographystyle{ieeetr}
\bibliography{a-ref}

\end{document}